\documentclass[twocolumn,english,pra, a4paper, twocolumn, 10pt]{revtex4}
\usepackage[T1]{fontenc}
\usepackage[latin9]{inputenc}
\usepackage{babel}
\usepackage{prettyref}
\usepackage{amsmath}
\usepackage{amssymb}
\usepackage{graphicx}
\usepackage{esint}
\usepackage[unicode=true,pdfusetitle,
 bookmarks=true,bookmarksnumbered=false,bookmarksopen=false,
 breaklinks=false,pdfborder={0 0 1},backref=false,colorlinks=false]
 {hyperref}

\makeatletter

\providecommand{\tabularnewline}{\\}

\@ifundefined{textcolor}{}
{%
 \definecolor{BLACK}{gray}{0}
 \definecolor{WHITE}{gray}{1}
 \definecolor{RED}{rgb}{1,0,0}
 \definecolor{GREEN}{rgb}{0,1,0}
 \definecolor{BLUE}{rgb}{0,0,1}
 \definecolor{CYAN}{cmyk}{1,0,0,0}
 \definecolor{MAGENTA}{cmyk}{0,1,0,0}
 \definecolor{YELLOW}{cmyk}{0,0,1,0}
}

\usepackage{ bbm }
\DeclareMathAlphabet{\mathbbold}{U}{bbold}{m}{n} 

\makeatother

\begin{document}

\title{Optimized Quantum Error Correction Codes for Experiments}

\author{V. Nebendahl}

\affiliation{Institut für Theoretische Physik, Universität Innsbruck, Technikerstr.
25, A-6020 Innsbruck, Austria}

\email{Volckmar.Nebendahl@uibk.ac.at}

\begin{abstract}
We identify gauge freedoms in quantum error correction (QEC) codes
and introduce strategies for optimal control algorithms to find the
gauges which allow the easiest experimental realization. Hereby, the
optimal gauge depends on the underlying physical system and the available
means to manipulate it. The final goal is to obtain optimal decompositions
of QEC codes into elementary operations which can be realized with
high experimental fidelities. In the first part of this paper, this
subject is studied in a general fashion, while in the second part,
a system of trapped ions is treated as a concrete example. A detailed
optimization algorithm is explained and various decompositions are
presented for the three qubit code, the five qubit code, and the seven
qubit Steane code.
\end{abstract}


\maketitle
\tableofcontents{}

\section{Introduction}

Similar as their classical counterparts, complex quantum algorithms
can be decomposed into sequences of elementary operations \cite{Barenco1995}.
These decompositions are not unique and depending on the purpose,
some of them might be better suited than others. In theoretical treatments
for instance, the elementary operations used for decompositions are
generally of such kind that they are as intuitive as possible. For
experimental realizations on the other hand, it is an obvious approach
to pick the elementary operations according to the available physical
processes in the quantum system of choice such that it can be manipulated
with the highest possible fidelity. As long as the elementary operations
allow us to assemble a universal set of quantum gates, the existence
of decompositions based on these operation is guaranteed for arbitrary
quantum algorithms \cite{Nielsen2000}. Still, in this paper we are
not just interested in the mere existence of such decompositions,
but seek ways to find decompositions which are as close to optimal
as possible. In particular, we focus on quantum error correction codes,
as well as operations on logical qubits encoded in error protected
states.

As explained in more detail below, these types of codes (respectively
algorithms) are equipped with different kinds of gauge invariances,
which offer greater freedom when we intend to decompose them. We are
no longer constrained to find a perfect one-to-one match. It suffices
to find a sequence of elementary operations which reproduces the given
code up to a gauge transformation. This increases the number of acceptable
solutions and with that the probability to find better sequences,
which allow high experimental fidelities.

To understand the origin of the quantum errpor correction (QEC) gauge
freedom we are interested in, we have to remind ourselves that errors
increase the entropy of a quantum system. QEC codes remove these errors
and thereby reduce the entropy. Since entropy can only be reduced
locally but not globally, one always needs to include a process which
maps the information of the error onto another system to balance the
entropy. This mapping is not unique and due to this lack of uniqueness,
each QEC code contains an inherent gauge freedom. Of course, these
mappings are subjected to constraints, but we are still left with
some degrees of freedom we can exploit.

For logical operations on error protected logical qubits, on the other
hand, the objective is that a correctable error is not amplified under
these logical operations, but it may still be mapped onto any other
correctable error. The freedom contained in this mapping can be exploited,
too.

Still, having identified these extra degrees of freedom does not imply
that we can harness them. Finding an optimal decomposition is a non
trivial task and it is advisable to seek the help of a (classical)
computer. Except for the gauge freedom, this is a typical task for
optimal control theory, which has already been applied successfully
to various different quantum systems \cite{Brif2010}. Therefore,
a central concern of this paper is to show how the above identified
gauge freedoms can be integrated in an optimal control algorithm.

Many of the ideas presented in this paper are of a general nature
and device independent, while on the other hand any specific decomposition
depends on the underlying quantum system. This dichotomy between generality
and peculiarity is also reflected by the setup of this paper. In the
first part, general aspects are treated which are common to all quantum
systems, while the results presented in Sec.~\ref{sec:Results} focus
on a specific system consisting of trapped ions \cite{Schindler2013}. 

In more detail, the structure of the paper is the following: In Sec.~\ref{sec:Quantum-error-correction},
the different kinds of gauge freedoms are worked out. Sec.~\ref{sec:Optimization}
outlines how these gauge freedoms translate into an optimal control
algorithm. In the Results section (Sec.~\ref{sec:Results}), various
decompositions are shown, comprising the three qubit code \cite{Shor1995},
the five qubit code \cite{Laflamme1996}, and the seven qubit Steane
code \cite{Steane1996}. The algorithm used to obtain these results
is explained in detail in Appendix~\ref{sec:Optimization-algorithm-for-Trapped ions}).

\section{Quantum error correction\label{sec:Quantum-error-correction}}

In this paper, we are mainly interested in experimental realizable
QEC codes, which favor simple approaches. Hence, the needed theoretical
background is quite limited and can be found, e.g.,\ in Ref.~\cite{Nielsen2000}. 

The QEC codes we are interested in all define two codewords, $|0_{L}\rangle$
and $|1_{L}\rangle$, to which we also refer as the logical states.
With them, an error protected logical qubit state $|\psi_{L}\rangle$
can be defined as
\begin{equation}
|\psi_{L}\rangle=\alpha\cdot|0_{L}\rangle+\beta\cdot|1_{L}\rangle\quad\textrm{with}\quad|\alpha|^{2}+|\beta|^{2}=1.\label{eq:Error Protected Qubit}
\end{equation}
The logical states $|0_{L}\rangle$ and $|1_{L}\rangle$ describe
two states in a higher-dimensional Hilbert space, which is given as
the product space of several single qubits, to which we refer as the
physical qubits.

The logical states are chosen such that even after the occurrence
of certain errors $E_{j}$, the logical qubit state $|\psi_{L}\rangle$
\eqref{eq:Error Protected Qubit} can still be reconstructed by an
error correcting code
\begin{equation}
E_{j}\cdot|\psi_{L}\rangle\overset{\textrm{QEC}}{\longrightarrow}|\psi_{L}\rangle.\label{eq:Schema QEC}
\end{equation}
Different QEC codes define different codewords for the logical states
$|0_{L}\rangle$ and $|1_{L}\rangle$ and allow for different errors
to be corrected. Further, the choice of the codewords influences the
way logical gates $U_{L}$, which operate on logical qubits $|\psi_{L}\rangle$,
have to be realized on the level of the physical qubits. 

In this paper, we are interested in finding the simplest possible
experimental realization of logical gates and quantum error correction
based on given codewords $|0_{L}\rangle$ and $|1_{L}\rangle$. That
is, our aim is not to invent new QEC codes with new codewords, but
to find variations of already existing codes.

\subsection{Degrees of freedom in QEC}

\begin{figure}
\includegraphics[width=1\columnwidth]{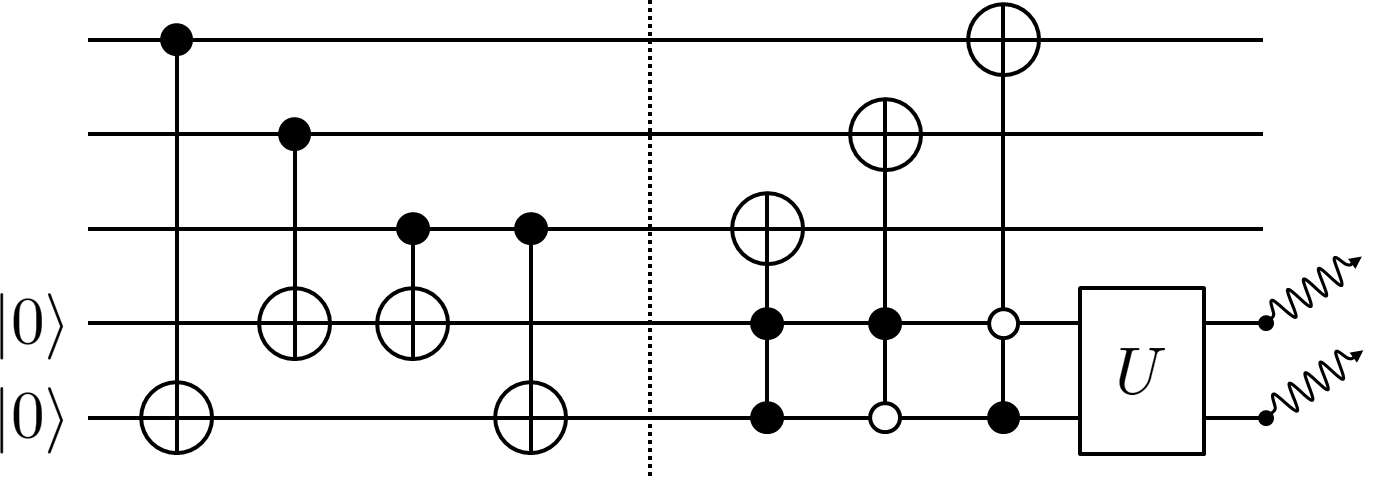}\protect\caption{\label{fig:Three Qubit Quantum-error-correction}Example circuit for
the three qubit quantum error correction code without measurement.
The top three horizontal lines represent the physical qubits used
to encode the logical qubit, while the two horizontal lines at the
bottom represent two auxiliary qubits. The dotted vertical line was
included to separate the two functional parts of the code: The part
to the left of the dotted line maps the information of a possible
bit-flip error onto the auxiliary qubits. The part on the righjt uses
this information to correct the error. After the error is corrected,
the state of the auxiliary qubits is of no further interest and might
be reset (wavy line) for the next run. Therefore, an arbitrary unitary
operation $U$ on the auxiliary system can be included.}
\end{figure}

Probably the simplest example for a QEC code is the three qubit code
\cite{Shor1995}, where the codewords for the logical qubit states
$|0_{L}\rangle$ and $|1_{L}\rangle$ consist of three physical qubits,
\begin{equation}
|0_{L}\rangle=|000\rangle\;\textrm{and}\;|1_{L}\rangle=|111\rangle.
\end{equation}
This encoding allows us to identify and correct a single bit-flip
error on any of the three physical qubits. A possible way to identify
such an error is to measure the physical qubits and use a majority
vote, but these measurements would also destroy the quantum information
stored in the logical qubit $|\psi_{L}\rangle$ \eqref{eq:Error Protected Qubit}.
This problem can be avoided if the information about the error is
mapped onto auxiliary qubits. Subsequently, the information in the
auxiliary qubits (the error syndrome) can either be measured or used
coherently to correct the error. Fig.~\ref{fig:Three Qubit Quantum-error-correction}
shows a possible circuit for a coherent error correction. After the
error is corrected, the logical qubit is once again disentangled from
the auxiliary qubits, whose final state is of no further interest.
In Fig.~\ref{fig:Three Qubit Quantum-error-correction}, this is
symbolized by the arbitrary unitary gate $U$ operating on the auxiliary
qubits at the end of the circuit. This unitary gate $U$ can be seen
as a gauge transformation \--- it changes the code, but it does not
change the result of the error correction on the logical qubit. Due
to this gauge freedom, we do not just have one bit-flip error correcting
QEC code but an entire equivalence class. 

Of course, simply adding such a unitary operator at the end of an
already perfectly working QEC is of no advantage at all. But we have
to remember the task we are facing: We are not looking for a decomposition
of the QEC code in terms of the standard gates used in Fig.~\ref{fig:Three Qubit Quantum-error-correction},
but for the simplest decomposition in operations which are tailored
for the quantum system we are using. In this search, it is a great
difference whether we have just one solution which we are allowed
to accept or an entire equivalence class of correct solutions.

\subsubsection{Measuring the error syndrome\label{sub:Measuring-the-error syndrom}}

\begin{figure}
\includegraphics[width=0.83\columnwidth]{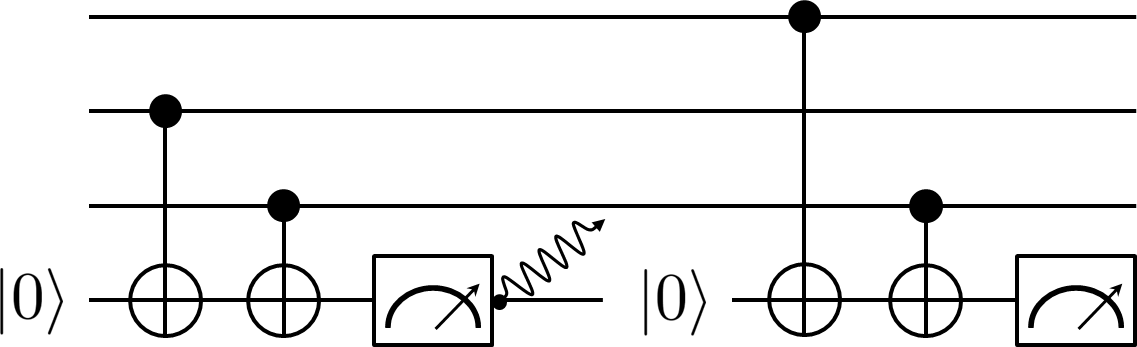}\protect\caption{\label{fig:Three Qubit QEC gemessen}Example circuit for a three qubit
quantum error correction code based on measurement. As in Fig.~\ref{fig:Three Qubit Quantum-error-correction},
the top three lines represent the physical qubits used to encode the
logical qubit, while the bottom line represents an auxiliary qubit.
After the first measurement, the auxiliary qubit is reset to reuse
it for the second measurement. The measurement results $(00)$, $(01)$,
$(10)$, and $(11)$ correspond to the four cases: no error, bit-flip
on the first qubit, bit-flip on the second qubit and bit-flip on the
third qubit.}
\end{figure}

Now, we turn to QEC codes which incorporate measurements of the auxiliary
qubits to determine the error syndrome. In the case of the three qubit
code example a bit-flip error can be identified using the quantum
circuit shown in Fig.~\ref{fig:Three Qubit QEC gemessen}. The possible
measurement outcomes $(00)$, $(01)$, $(10)$, and $(11)$ correspond
to the four cases: no error, bit-flip on the first physical qubit,
bit-flip on the second physical qubit, and bit-flip on the third physical
qubit. Here, we face a fixed mapping between the measurement results
and the error. Hence, the situation is different compared to the measurement-free
example above, where we did not care for the final state of the auxiliary
qubits. 

Nonetheless, we can still consider alternative mappings. But if we
do so, we have to make sure that these mappings are unambiguous, i.e.,\ it
is not enough that the different measurement results correspond to
orthogonal error states, but one also has to avoid measurement results
which correspond to superpositions of errors, since otherwise the
correction operation would be ill defined. As long as we take care
of this, we can even allow codes which change the error on the logical
qubit, as long as the error maintains correctable. 

Such alternative mappings cause no problems if we choose them by hand.
But we are looking for degrees of freedom which can be manipulated
by the computer itself in the attempt to find optimal decompositions.
If we try to let the computer find the best mapping, we have to make
sure that the above-mentioned side conditions are fulfilled, which
causes the optimization problem to become much more involved. A further
difficulty arises for optimization problems which are so complex that
we have to split them into several smaller problems (e.g.,\ stabilizer
codes, where the code for each stabilizer might be computed separately;
see Sec.~\prettyref{sub:Stabilizer}). If we allow the computer to
alter the mapping in such a case, this change has to be synchronized
for all subprocesses. Since these subprocesses can no longer be handled
independently, splitting a complex problem into smaller subprocesses
might not result in the intended simplification of the problem at
all. But there are also some mappings which entail none of the above
mentioned problems. These mappings, we are going to study next.

First, let us define the following symbols
\begin{eqnarray}
E_{j} &  & \textrm{elementary correctable error}\nonumber \\
|0_{L}\rangle,|1_{L}\rangle &  & \textrm{basis of the logical qubit}\nonumber \\
|0_{A}\rangle &  & \textrm{inital state of the auxiliary qubits}\nonumber \\
|e_{j}\rangle &  & \begin{array}{l}
\textrm{final state of the auxiliary qubits }\\
\textrm{corresponding to the error }E_{j}.
\end{array}\label{eq:Buchstaben for QEC}
\end{eqnarray}
The $E_{j}$ form a minimal basis of operators for all correctable
errors (i.e.~$\alpha\cdot E_{i}+\beta\cdot E_{j}$ is a correctable
error while $E_{i}\cdot E_{j}$ is generally not), including the identity.
Each $E_{j}$ corresponds to exactly one possible measurement result
$|e_{j}\rangle$ of the auxiliary qubits. 

To identify the error, we need a unitary operator $U_{\textrm{Syn.}}$
which calculates the error syndrome, i.e.\ $U_{\textrm{Syn.}}$ maps
the error $E_{j}$ onto the auxiliary qubits $|0_{A}\rangle\rightarrow|e_{j}\rangle$,
while the logical qubit stays unchanged. The most general form for
such an operator $U_{\textrm{Syn.}}$ is given by

\begin{align}
U_{\textrm{Syn.}} & =\sum_{j}e^{i\varphi_{j}}E_{j}\left(|0_{L}\rangle\langle0_{L}|+|1_{L}\rangle\langle1_{L}|\right)E_{j}^{\dagger}\otimes|e_{j}\rangle\langle0_{A}|\nonumber \\
 & +M_{C}\label{eq:U syndrom}
\end{align}
with arbitrary phases $\varphi_{j}$. Here, $|0_{L}\rangle\langle0_{L}|+|1_{L}\rangle\langle1_{L}|$
is the identity in the Hilbert space of the logical qubits, but it
is \emph{not }an identity in the Hilbert space of the physical qubits,
where the error occurs. Further, we added $M_{C}$ in Eq.~\eqref{eq:U syndrom}
for formal reasons only, i.e.\ to complement $U_{\textrm{Syn.}}$
to a unitary operator. $M_{C}$ operates exclusively on states which
are in the kernel of the operator sum in the first line of Eq.~\eqref{eq:U syndrom},
as, e.g.,\ input states where the auxiliary qubits are not initialized.
Apart from the unitarity constraint of $U_{\textrm{Syn.}}$, the operator
$M_{C}$ is completely arbitrary. 

In textbook decompositions of $U_{\textrm{Syn.}}$ \eqref{eq:U syndrom},
the phase factors $e^{i\varphi_{j}}$ are usually set to $e^{i\varphi_{j}}=1$,
but $\varphi_{j}$ can also adopt any other value, without effecting
the result of the error correction. The phase factors $e^{i\varphi_{j}}$
introduce relative phases between the superposition branches belonging
to different elementary errors. Since all QEC codes are designed such
that only one of these elementary errors survives after the measurements
\cite{Nielsen2000}, the remaining phase solely adds to the global
phase and hence does no harm %
\footnote{More complex codes might also be able to handle two incorrect qubits,
but this still counts as one error in the sense that we have to define
an elementary error $E_{j}$ which corresponds to the combination
of two corrupted qubits. Hence, it is still just one phase factor
that remains.%
}.

\subsection{Fault-tolerance\label{sub:Fault-tolerance}}

The optimization algorithm we will derive in this paper is designed
to find a suitable sequence of technically feasible elementary operations
such that the collective action of this sequence is identical to the
effect of a given QEC code. For practical reasons, this is all we
will demand, but from the theoretical point of view, one might demand
more. To comply with the paradigm of the fault-tolerant quantum computation
\cite{Nielsen2000,DiVincenzoShor1996,Threshold_Preskill1998}, we
also would have to ensure that a single error occurring during the
execution of the sequence does not result in the total loss of a formerly
error-free logical qubit (or at least, the probability of a total
loss has to be sufficiently small). Evidently, including such an extra
demand would complicate the optimization task, but it would also impose
constraints upon the selection of the elementary operations. For example,
it is hard to conceive a fault-tolerant sequence based on elementary
operations which corrupt many qubits at once in the case of a single
error. In total, the demand for fault tolerance might hinder us to
find any practical solution at all. Thus, we adopt a pragmatical point
of view: As long as the experimental capacities to realize quantum
error correction are still far behind the theoretical visions, it
seems more important to provide experimentally realizable code than
to have all theoretical aspects covered. Therefore, we feel safe to
ignore some demands of fault tolerance for the time being and postpone
their solution into the future. 

Nonetheless, we will not sacrifice all ideas of fault tolerance. Beside
the ability to correct errors, we also need to be able to apply logical
operators (quantum gates) on logical qubits. We still demand that
correctable errors (which are already present before the logical operators
are applied) are not amplifies by these logical operators.

\subsection{Operations on logical qubits\label{sub:Operations-on-logical qubits}}

Here, we restrict ourselves to logical operators which act on a single
logical qubit only. In order to avoid error amplification, we could
demand that any logical operator $\hat{O}$ has to commutate with
any correctable error $E_{j}$, when applied to an arbitrary logical
qubit $|\psi_{L}\rangle$, i.e., $\hat{O}\cdot E_{j}|\psi_{L}\rangle=E_{j}\cdot\hat{O}|\psi_{L}\rangle$.
But this is more than needed. It suffices that under any logical operation
$\hat{O}$ correctable errors are only allowed to transform into other
correctable errors, i.e., 
\begin{equation}
\hat{O}\cdot E_{j}|\psi_{L}\rangle=\sum_{k}\alpha_{jk}\cdot E_{j}\cdot\hat{O}|\psi_{L}\rangle,\quad\sum_{k}|\alpha_{jk}|^{2}=1,\label{eq:Maximally Free Kommutator}
\end{equation}
with the extra degree of freedom to adopt the $\alpha_{jk}$.

On the other hand, we have to remember that the identity operation
is part of the elementary errors $\{E_{j}\}$, as well. That is, although
Eq.\ \eqref{eq:Maximally Free Kommutator} avoids error amplification,
it allows that an error-free state is mapped onto an incorrect (but
still correctable) result. To prevent this from happening, it is advisable
to introduce a hierarchy of correctable errors and to distinguish
at least the hierarchy levels ``zero incorrect qubits'' and ``one
incorrect qubit.'' For more advanced codes, further levels might
be added. With that, we demand that errors are only mapped onto errors
of the same hierarchy level. Using the index $h$ to enumerate the
hierarchy levels and denoting the errors by $E_{h,j_{h}}$, this demand
can be formulated as 
\begin{equation}
\hat{O}\cdot E_{h,j_{h}}|l\rangle=\sum_{k_{h}}\alpha_{j_{h}k_{h}}\cdot E_{h,k_{h}}\hat{O}|\psi_{L}\rangle,\;\sum_{k_{h}}|\alpha_{j_{h}k_{h}}|^{2}=1.
\end{equation}
At a first glance, one might also want to accept the case where some
errors are reduced due the application of the operator $\hat{O}$,
but this is ruled out by entropy if we do not allow error accumulation
elsewhere for compensation.

\section{Performance function\label{sec:Optimization}}

After having introduced the different degrees of freedom we intend
to use, we have to investigate how they are integrated into the optimization
routine.

We like to optimize QEC codes in the circuit model of quantum computation.
These codes might consist of several components, as, e.g.,\ unitary
operations which map an error syndrome onto some auxiliary qubits
followed by a measurement and finally an active correction of the
error according to the measurement results \cite{Nielsen2000}. The
measurement and the active correction are usually elementary operations
and considered as trivial in this context. The only part we intend
to optimize are the complex unitary operations. 

Any unitary operator $U$ has to be built up using some elementary
operations $u_{t}(\alpha_{t})$ which can be generated in the laboratory,
\begin{equation}
U=U(\boldsymbol{\alpha})=\prod_{t}u_{t}(\alpha_{t}).\label{eq:Prodult_Unitar}
\end{equation}
The origin of this decomposition \eqref{eq:Prodult_Unitar} depends
on the treatment of the quantum system in question. The $u_{t}(\alpha_{t})$
might represent some effective unitaries which can be generated easily
and with high fidelity. This is, e.g.,\ the case for the second part
of this paper, where we demonstrate the ideas for trapped ions. Alternatively,
the $u_{t}(\alpha_{t})$ might result from the necessity to discretize
the general time ordered equation
\begin{equation}
U=\mathcal{T}\exp\left(-\frac{i}{\hbar}\int_{0}^{T}dt\cdot H\left(\alpha(t)\right)\right)
\end{equation}
to treat it numerically. Depending on the situation, each $\alpha_{t}$
either stands for a single control parameter or represents an entire
set of such parameters, $\alpha_{t}\equiv\alpha_{t,j}$. In either
case, these are the parameters we have to optimize. 

How is this optimization done? Let us start with a simple example.
Suppose we like to generate the operator $U_{\textrm{target}}$. As
a first step, we define the real valued performance function $\Phi_{\textrm{Re}}(\boldsymbol{\alpha})$,
\begin{eqnarray}
\Phi_{\textrm{Re}}(\boldsymbol{\alpha}) & := & \textrm{Re}(\langle U_{\textrm{target}}|\prod_{t}u_{t}(\alpha_{t})\rangle)\nonumber \\
 & = & \textrm{Re}(\textrm{tr}(U_{\textrm{target}}^{\dagger}\cdot\prod_{t}u_{t}(\alpha_{t}))).\label{eq: Gain function example 1}
\end{eqnarray}
 This function reaches its maximum if and only if 
\begin{equation}
\prod u_{t}(\alpha_{t})=U_{\textrm{target}}.
\end{equation}
In principal, we could now use a standard optimization procedure as,
e.g.,\ a gradient ascent algorithm \cite{Khaneja2005,SchulteHerbruggen2011}
to find solutions for the $\alpha_{t}$. Therefore, we consider this
example problem as solved.

Next, suppose we take the absolute value in Eq.~\eqref{eq: Gain function example 1}
instead of the real value,
\begin{equation}
\Phi_{\textrm{Abs}}(\boldsymbol{\alpha}):=|\langle U_{\textrm{target}}|\prod_{t}u_{t}(\alpha_{t})\rangle|^{2}.\label{eq:Gain function example abs}
\end{equation}
This performance function is now maximized by all $\boldsymbol{\alpha}$
with
\begin{equation}
\prod u_{t}(\alpha_{t})=e^{i\varphi}\cdot U_{\textrm{target}}.
\end{equation}
In case we can tolerate the extra phase factor $\exp(i\varphi)$,
$\Phi_{\textrm{Abs}}(\boldsymbol{\alpha})$ \eqref{eq:Gain function example abs}
defines the preferable performance function since it exploits an extra
degree of freedom and accepts more solutions. Here, we get a first
idea how the degrees of freedom found for QEC codes in Sec.~\ref{sec:Quantum-error-correction}
might be used.

\subsection{Error syndrome}

Now, let us construct a performance function which is maximal if $\prod u_{t}(\alpha_{t})$
equals any of the $U_{\textrm{Syn.}}$ described in Eq.~\eqref{eq:U syndrom}.
It might be helpful to keep in mind that the performance function
is just a mathematical tool, which helps us to find the correct $\alpha_{t}$.
It does not need to have a meaningful physical interpretation.

First, we make the obvious choice not to include the unimportant part
$M_{C}$ \eqref{eq:U syndrom} into the performance function. $M_{C}$
acts exclusively on states we are not interested in and was only introduced
for the formal reason to make $U_{\textrm{Syn.}}$ unitary. Dropping
it reduces the class of adequate unitaries $U_{\textrm{Syn.}}$ to
a class of projectors with elements $P_{\textrm{Syn.}}$,
\begin{align}
P_{\textrm{Syn.}} & =\sum_{j}e^{i\varphi_{j}}E_{j}\left(|0_{L}\rangle\langle0_{L}|+|1_{L}\rangle\langle1_{L}|\right)E_{j}^{\dagger}\otimes|e_{j}\rangle\langle0_{A}|\nonumber \\
 & =\sum_{j}\sum_{l=1}^{2}e^{i\varphi_{j}}\cdot E_{j}|l_{L}\rangle\langle l_{L}|E_{j}^{\dagger}\otimes|e_{j}\rangle\langle0_{A}|
\end{align}
Next, we introduce the double indexed object $p_{lj}$ as
\begin{equation}
p_{lj}=E_{j}|l_{L}\rangle\langle l_{L}|E_{j}^{\dagger}\otimes|e_{j}\rangle\langle0_{A}|,\quad l=0,1.\label{eq:Objekt p_lj syndrom}
\end{equation}
As can be easily checked, $P_{\textrm{Syn.}}=\sum_{lj}e^{i\varphi_{j}}\cdot p_{lj}$. 

The $p_{lj}$ are fixed and known objects with no further degree of
freedom. To avoid confusions, we emphasize that the objects $p_{lj}$
are operators, but contrary to common notations the indices $l,j$
do not enumerate bra and ket bases, i.e.,\ $\hat{p}\neq p_{lj}\cdot|l\rangle\langle j|$.

Now, as a first (and insufficient) attempt, one might try the performance
function 
\begin{eqnarray}
\Phi_{\textrm{try}}(\boldsymbol{\alpha}) & = & \sum_{lj}\bigl|\langle p_{lj}|\prod_{t}u_{t}(\alpha_{t})\rangle\bigr|^{2}\nonumber \\
 & = & \sum_{lj}\bigl|\textrm{tr}(p_{lj}^{\dagger}\cdot\prod_{t}u_{t}(\alpha_{t}))\bigr|^{2}.
\end{eqnarray}
This performance function is maximal, if all summands $|\textrm{tr}(p_{lj}^{\dagger}\cdot\prod_{t}u_{t}(\alpha_{t}))|$
are maximal. That is, $\Phi_{\textrm{try}}(\boldsymbol{\alpha})$
is maximized, if 
\begin{equation}
\textrm{tr}(p_{lj}^{\dagger}\cdot\prod_{t}u_{t}(\alpha_{t}))=e^{i\varphi_{lj}},
\end{equation}
for all $l,j$. The problem with this performance function is that
the phases $e^{i\varphi_{lj}}$ for the logical states $|0_{L}\rangle$
and $|1_{L}\rangle$ are not synchronized, i.e.\ the performance
function is maximized for all
\begin{equation}
\prod_{t}u_{t}(\alpha_{t})=\sum_{lj}e^{i\varphi_{lj}}\cdot E_{j}|l_{L}\rangle\langle l_{L}|E_{j}^{\dagger}\otimes|e_{j}\rangle\langle0_{A}|+M_{C},
\end{equation}
with phases $\varphi_{lj}$ that depend on $|l\rangle$. This can
be easily amended by summing over the index $l$ \emph{before} taking
the absolute value,
\begin{equation}
\Phi_{\textrm{two sums}}(\boldsymbol{\alpha})=\sum_{j}\Bigl|\sum_{l}\langle p_{lj}|\prod_{t}u_{t}(\alpha_{t})\rangle\Bigr|^{2}.
\end{equation}
The absolute square of the sum over $l=0,1$ generates three terms:
$|\langle p_{0,j}|\prod_{t}u_{t}(\alpha_{t})\rangle|^{2}$, $|\langle p_{1,j}|\prod_{t}u_{t}(\alpha_{t})\rangle|^{2}$,
and $\textrm{Re}\bigl(\langle p_{0,j}|\prod_{t}u_{t}(\alpha_{t})\rangle\cdot\langle\prod_{t}u_{t}(\alpha_{t})|p_{1,j}\rangle\bigr)$.
A closer look reveals that it suffices to take only the last term
into account, i.e.\ the final version of the performance function
for the error syndrome becomes 
\begin{equation}
\Phi(\boldsymbol{\alpha})=\sum_{j}\textrm{Re}\Bigl(\langle p_{0,j}|\prod_{t}u_{t}(\alpha_{t})\rangle\cdot\langle\prod_{t}u_{t}(\alpha_{t})|p_{1,j}\rangle\Bigr),\label{eq:Gain function Syndrom}
\end{equation}
which can only become maximal if $\langle p_{0,j}|\prod_{t}u_{t}(\alpha_{t})\rangle=e^{i\varphi_{j}}=\langle p_{1,j}|\prod_{t}u_{t}(\alpha_{t})\rangle$
(for each $j$).

By now, we have found a performance function $\Phi(\boldsymbol{\alpha})$
which allows us to optimize over all degrees of freedom we have identified
for the unitary operation that maps the error onto the auxiliary qubits.
The same approach can be used if we like to map a stabilizer \cite{Nielsen2000,Gottesman1996}
onto an auxiliary qubit. For the coherent QEC approach without measurement
(see the beginning of Sec.~\ref{sec:Quantum-error-correction}) on
the other hand, we need a new performance function. This is what we
are constructing next.

\subsection{Coherent QEC}

First, we define a new double indexed object $\tilde{p}_{lj}$ as
\begin{equation}
\tilde{p}_{lj}=|l_{L}\rangle\langle l_{L}|E_{j}^{\dagger}\otimes\langle0_{A}|,\quad l=0,1.\label{eq:p-object fully quantum}
\end{equation}
The input states (right or ``bra'' side) are the same as for $p_{lj}$
\eqref{eq:Objekt p_lj syndrom} but the output states are different.
Since the coherent QEC corrects the error, we find the error-free
states $|l_{L}\rangle$ instead of $E_{j}|l_{L}\rangle$. Further,
no output is determined into the Hilbert space of the auxiliary qubits,
which corresponds to the freedom we have identified for this model.
To see this more clearly, we have to look at 
\begin{eqnarray}
\langle\tilde{p}_{lj}|\prod_{t}u_{t}(\alpha_{t})\rangle & = & \textrm{tr}(\tilde{p}_{lj}^{\dagger}\cdot\prod_{t}u_{t}(\alpha_{t}))\nonumber \\
 & = & (\langle l_{L}|\cdot\prod_{t}u_{t}(\alpha_{t})\cdot E_{j}|l_{L}\rangle\otimes|0_{A}\rangle)\nonumber \\
 & = & |a_{lj}\rangle.
\end{eqnarray}
For fixed $l,j$, the ket vector $|a_{lj}\rangle$ describes a state
in the Hilbert space of the auxiliary qubits with $\Vert|a_{lj}\rangle\Vert\leq1$.
The value $\Vert|a_{lj}\rangle\Vert=1$ is only adopted iff the erroneous
state $E_{j}|l_{L}\rangle$ is corrected to $|l_{L}\rangle$ by $\prod_{t}u_{t}(\alpha_{t})$.
Let us recapitulate what we know about the physical state of the auxiliary
qubits after a successful coherent error correction: The state is
allowed to be arbitrary, but of course it has to be normalized to
one and the state has to be independent of the logical state $|l_{L}\rangle$,
since otherwise the auxiliary qubits would be entangled with the logical
qubit. All these conditions are met if the expression 
\begin{equation}
\sum_{j}\textrm{Re}(\langle a_{0,j}|a_{1,j}\rangle)
\end{equation}
is maximized. With that, we obtain the performance function for the
coherent QEC approach,
\begin{equation}
\tilde{\Phi}(\boldsymbol{\alpha})=\sum_{j}\textrm{Re}\Bigl(\langle\tilde{p}_{0,j}|\prod_{t}u_{t}(\alpha_{t})\rangle\cdot\langle\prod_{t}u_{t}(\alpha_{t})|\tilde{p}_{1,j}\rangle\Bigr).\label{eq:Gain Function fully quantum}
\end{equation}

\subsection{Quantum gates on logical qubits}

Finally, we turn to the performance function for quantum gates on
logical qubits. In Sec.~\ref{sub:Operations-on-logical qubits} we
demanded that any quantum gate $\hat{O}$ should map errors onto errors
of the same level of hierarchy. The needed projectors for such mappings
are given by 
\begin{equation}
\hat{p}_{l,h,j_{h},k_{h}}=E_{h,k_{h}}\cdot\hat{O}|l_{L}\rangle\langle l_{L}|E_{h,j_{h}}^{\dagger},
\end{equation}
where $h$ denotes the hierarchy level. Except for the extra indices,
the new performance function can be formulated with $\hat{p}_{h,j_{h},k_{h},l}$
using the same structure as for the two examples seen before \eqref{eq:Gain function Syndrom}
and \eqref{eq:Gain Function fully quantum}
\begin{equation}
\hat{\Phi}(\boldsymbol{\alpha})=\sum_{h,j_{h},k_{h}}\textrm{Re}\Bigl(\langle\hat{p}_{0,h,j_{h},k_{h}}|U(\boldsymbol{\alpha})\rangle\cdot\langle U(\boldsymbol{\alpha})|\hat{p}_{1,h,j_{h},k_{h}}\rangle\Bigr),
\end{equation}
with $U(\boldsymbol{\alpha})=\prod_{t}u_{t}(\alpha_{t})$. It might
be helpful to compare this performance function to a version without
the summation over $k$ (i.e., $k=j).$ Such a performance function
is deprived of the freedom to alter the type of error. Still, due
to the fact that $U(\boldsymbol{\alpha})$ is unitary, the maximal
value of both performance functions is exactly the same \--- it is
just not reached for so many $U(\boldsymbol{\alpha})$ if the summation
over $k$ is missing.

Remark: For the coherent QEC we omitted the ket basis for the auxiliary
qubits in $\tilde{p}_{lj}$ \eqref{eq:p-object fully quantum}. Alternatively,
we could also have included a $k$-indexed basis followed by a summation
over this index in the performance function.

\subsection{Evaluating the performance function\label{sub:Evaluating-the-performance}}

In order to optimize the performance function, one needs to vary over
the parameters $\alpha_{t}$. This can be done in an efficient manner
\cite{Khaneja2005}, as we are about to see for the derivatives $(\frac{\partial}{\partial\alpha_{0}},\frac{\partial}{\partial\alpha_{1}},\dots,\frac{\partial}{\partial\alpha_{T}})$.
To keep the demonstration simple, we treat each $\alpha_{t}$ as a
single parameter, although it might also represent an entire set of
parameters (see above). 

For further convenience, we write $p_{lj}$ \eqref{eq:Objekt p_lj syndrom}
as
\begin{equation}
p_{lj}=|f_{lj}\rangle\langle i_{lj}|.
\end{equation}
 With that, the central element of the performance function \eqref{eq:Gain function Syndrom}
reads
\begin{eqnarray}
\langle p_{lj}|\prod_{t=0}^{T}u_{t}(\alpha_{t})\rangle & = & \textrm{tr}\langle p_{lj}^{\dagger}\cdot\prod_{t=0}^{T}u_{t}(\alpha_{t})\rangle\nonumber \\
 & = & \langle f_{lj}|\prod_{t=0}^{T}u_{t}(\alpha_{t})|i_{lj}\rangle
\end{eqnarray}
and its derivatives with respect to $\alpha_{k}$ become 
\begin{equation}
\frac{\partial}{\partial\alpha_{k}}\langle\dots\rangle=\langle f_{lj}|\prod_{t=0}^{k-1}u_{t}(\alpha_{t})\cdot\frac{du_{k}(\alpha_{k})}{d\alpha_{k}}\cdot\prod_{t=k+1}^{T}u_{t}(\alpha_{t})|i_{lj}\rangle.\label{eq:Produkt aufspaltung}
\end{equation}
Since $\langle f_{lj}|\prod_{t=0}^{k-1}u_{t}(\alpha_{t})$ and $\prod_{t=k+1}^{T}u_{t}(\alpha_{t})|i_{lj}\rangle$
can be calculated iteratively for all $k$, the number of matrix multiplications
needed to calculate \emph{all} derivatives $(\frac{\partial}{\partial\alpha_{0}},\dots,\frac{\partial}{\partial\alpha_{T}})$
scales down from quadratic in $n$ to linear in $n$, which allows
e.g.\ an efficient application of a gradient ascent algorithm \cite{Khaneja2005,SchulteHerbruggen2011}.

\section{Results\label{sec:Results}}

So far, the discussion was fairly general and independent of any specifications
due to the chosen physical system. But these specifications are important
to generate an optimal code. Therefore, we now make a choice and show
the optimization results for a quantum system consisting of trapped
ions \cite{Schindler2013}. Still, we do not need to delve deep into
the physics of trapped ions. The only system dependent specification
we need is a description of the elementary operations $u_{t}(\alpha_{t})$
appearing in Eq.~\eqref{eq:Prodult_Unitar}. 

For trapped ions, we use effective unitaries as elementary operations,
from which we know how to produce them with very high fidelities.
The Hilbert space these effective unitaries operate on is entirely
restricted to the Hilbert space of the qubits. The advantage of such
an effective approach is a much simpler optimization task. The effective
unitaries $u_{t}(\alpha_{t})$ over which we optimize can be described
by a set of effective $N$-qubit Hamiltonians $H_{j}$ as
\begin{eqnarray}
u_{t}(\alpha_{t}) & = & \exp\left(-i\cdot\alpha_{t}\cdot H_{j_{t}}\right),\quad\textrm{with}\nonumber \\
H_{j} & \in & \{S_{x}^{2},S_{y}^{2},S_{x},S_{y},\sigma_{z}^{(1)},\sigma_{z}^{(2)},\dots,\sigma_{z}^{(N)}\},\nonumber \\
S_{x/y} & =\frac{1}{2} & \sum_{j=1}^{N}\sigma_{x/y}^{(j)},\label{eq:Effective Ham}
\end{eqnarray}
where $\sigma_{x/y/z}^{(j)}$ denote Pauli matrices applied on the
$j$th qubit. Linear combinations of the Hamiltonians are not allowed,
which is e.g.\ in great contrast to the typical optimization task
found in nuclear magnetic resonance (NMR) \cite{Khaneja2005}. The
$\exp(-i\cdot\alpha_{t}\cdot S_{x/y}^{2})$ represent Mølmer-Sørensen
gates \cite{Moelmer1999}. They are the only entangling gates in this
set and were chosen because they are currently the entangling gates
with the highest fidelities \cite{Benhelm2008}. More motivations
for the specific choice of the $u_{t}(\alpha_{t})$ in Eq.~\eqref{eq:Effective Ham}
are given in App.~\ref{sec:Elementary-operations-for-Trapped-ions}.

To identify the effective unitaries $u_{t}(\alpha_{t})=\exp(-i\cdot\alpha_{t}\cdot H_{\textrm{eff}}^{(t)})$,
we need two pieces of information: $\alpha_{t}$ and $H_{\textrm{eff}}^{(t)}$.
To this purpose, we use the following notation:
\begin{eqnarray}
z_{j}(\varTheta) & = & \exp\bigl(-i\cdot\varTheta\cdot\frac{\sigma_{z}^{(j)}}{2}\bigr)\nonumber \\
X(\varTheta) & = & \exp(-i\cdot\varTheta\cdot S_{x})\nonumber \\
Y(\varTheta) & = & \exp(-i\cdot\varTheta\cdot S_{y})\nonumber \\
X^{2}(\varTheta) & = & \exp(-i\cdot\varTheta\cdot S_{x}^{2})\nonumber \\
Y^{2}(\varTheta) & = & \exp(-i\cdot\varTheta\cdot S_{y}^{2}),\label{eq:Puls Notation}
\end{eqnarray}
with $S_{x/y}=\frac{1}{2}\sum_{j=1}^{N}\sigma_{x/y}^{(j)}$. Further,
we introduce the two symbols
\begin{eqnarray}
M_{j} & : & \textrm{measurement of \ensuremath{j}}\textrm{th qubit}\nonumber \\
R_{j} & : & \textrm{reset of \ensuremath{j}}\textrm{th qubit to }|1\rangle.\label{eq:Appendix Puls Notation}
\end{eqnarray}

\begin{figure}
\includegraphics[width=0.676\columnwidth]{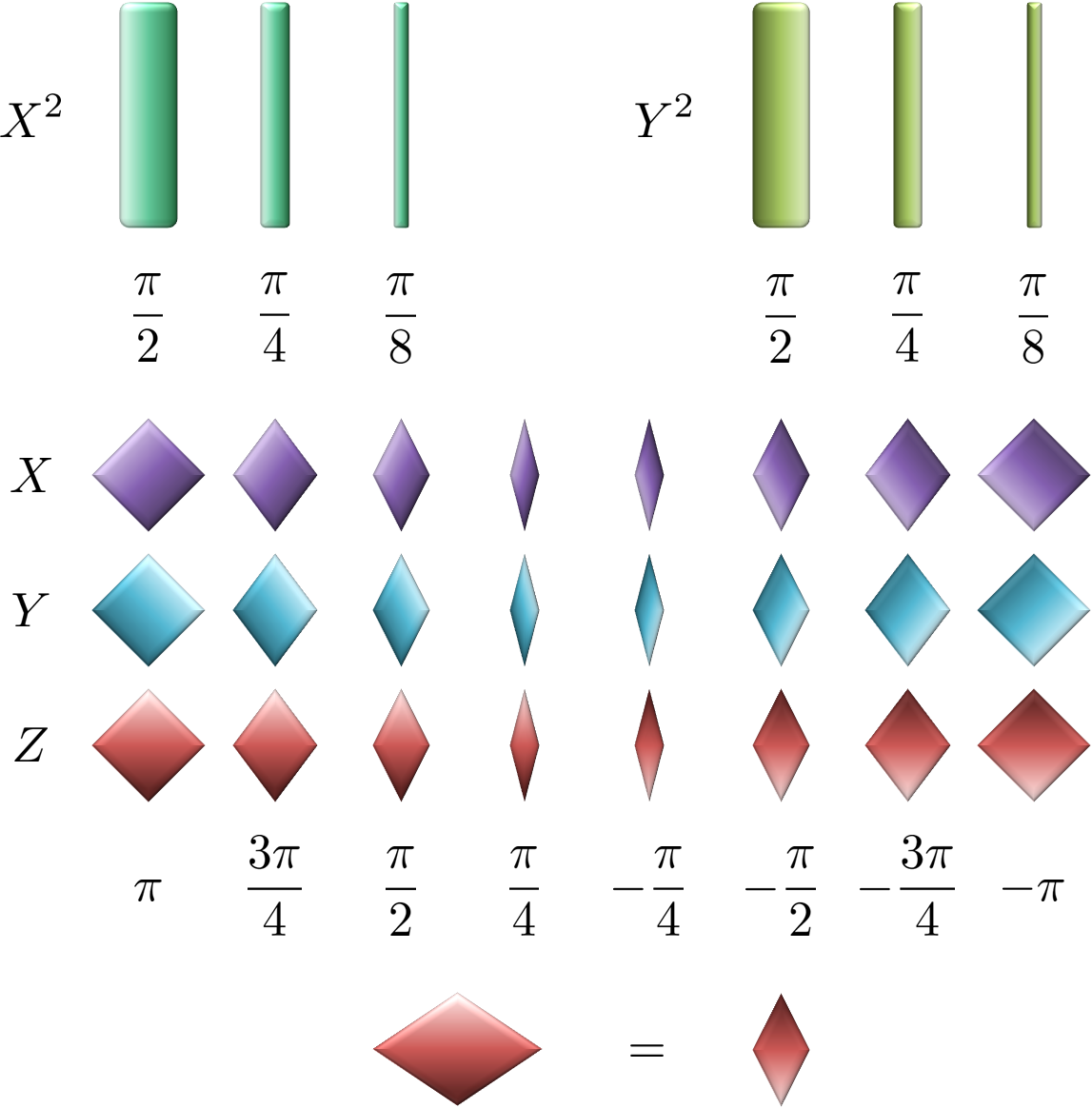}

\protect\caption{\label{fig:Symbols}The symbols used for the graphical representation
of $X(\varTheta),\ Y(\varTheta),\ z_{j}(\varTheta),\ X^{2}(\varTheta)$,
and $Y^{2}(\varTheta)$ \eqref{eq:Puls Notation} in a quantum circuit.
The width of the symbols is proportional to the absolute value of
$\varTheta$, where the width belonging to $\varTheta=\pi$ equals
the vertical distance of the qubits in the quantum circuit. The symbols
for the entangling Mølmer-Sørensen gates $X^{2}(\varTheta)$ and $Y^{2}(\varTheta)$
are extended over all qubits, while the symbols for the non-entangling
collective operations $X(\varTheta)$ and $Y(\varTheta)$ are (vertically)
repeated such that all qubits are covered. The symbol for the local
$z_{j}(\varTheta)$ is placed on the $j$th qubit only. Currently,
$z_{j}(\varTheta)$ with negative values for $\varTheta$ are experimentally
realized as $z_{j}(-|\varTheta|)=z_{j}(2\pi-|\varTheta|)$. Therefore,
the symbols for negative $z_{j}$ are replaced by their longer positive
counterparts, as well.}
\end{figure}

Besides the mathematical transcription, each result is also given
as graphical representation. Here, we follow the conventions used
for quantum circuit: Qubits are depicted as horizontal lines and operators
(gates) are given by symbols placed on the lines of the qubits in
question. The symbols used for the operators are explained in Fig.~\ref{fig:Symbols}.
Further, time order is from left to right. To comply with the graphical
representation, we also adapt the left to right time ordering for
the mathematical transcription, i.e.\ we write $X(\pi)\ z_{3}(\frac{\pi}{2})$
when $X(\pi)$ is applied first and $z_{3}(\frac{\pi}{2})$ second.

In this section, we present results for the three qubit code, the
five qubit code and the Steane code. In a typical realization, the
ground state of the ions is used to represent the qubit state $|1\rangle$.
Therefore, we also use the $|1\rangle$ as initial value for all qubits,
contrary to the more common $|0\rangle$ state, which we used e.g.,\ in
Figs.~\ref{fig:Three Qubit Quantum-error-correction} and \ref{fig:Three Qubit QEC gemessen}.

Readers who are interested in the details of the algorithm used to
obtain the results of this section find a description in App.~\ref{sec:Optimization-algorithm-for-Trapped ions}.

\subsection{Three qubit code}

Here, we treat the three qubit bit-flip and three qubit phase-flip
error correction codes \cite{Shor1995}. Both are realized on $3+1$
qubits, where the first three physical qubits encode the logical qubit
and the fourth qubit is an auxiliary qubit initialized to the $|1\rangle$
state.

\begin{figure}
\includegraphics[width=1\columnwidth]{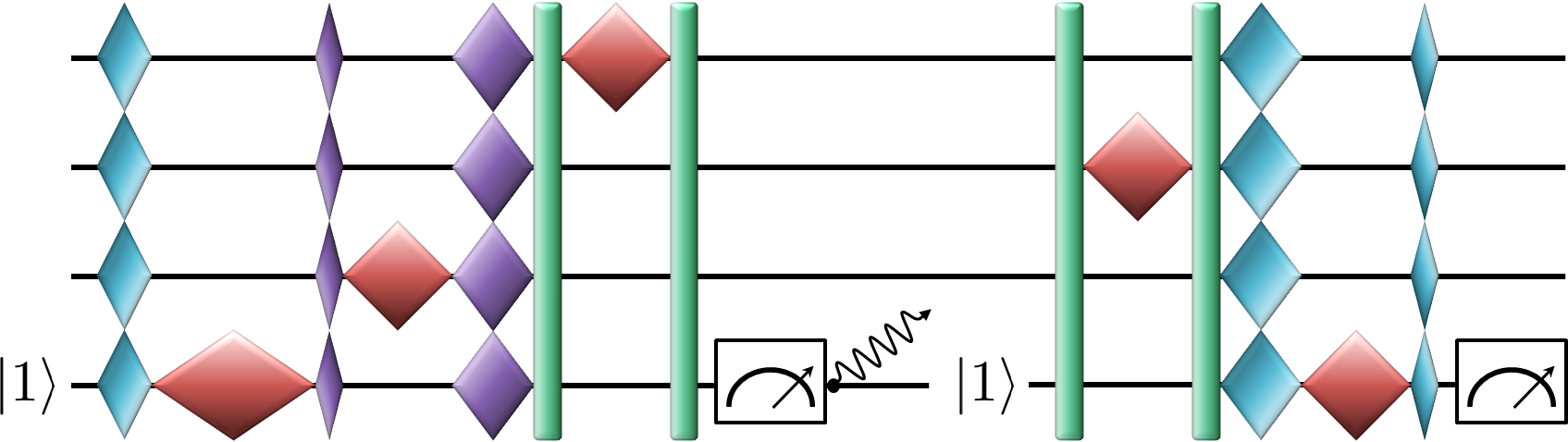}

\protect\caption{\label{fig:Three-qubit-bit-flip messen}Measurement of the error syndrome
for the three qubit bit-flip error correction code \eqref{eq:Sequenz 3 Qubit BitFlip Messen}.
The first three qubits encode the logical qubit. After the first measurement,
the auxiliary qubit is reset (wavy line). }
\end{figure}

We start with the readout of the error syndrome for the bit-flip error
correction code (equivalent to the quantum circuit shown in Fig.~\ref{fig:Three Qubit QEC gemessen}).
Here, the codewords for the logical qubit are $|0_{L}\rangle=|000\rangle$,
$|1_{L}\rangle=|111\rangle$. The following sequence maps the error
syndrome onto the auxiliary qubit 
\begin{equation}
\begin{array}{lllll}
Y\left(-\frac{\pi}{2}\right) & z_{4}\left(-\frac{\pi}{2}\right) & X\left(-\frac{\pi}{4}\right) & z_{3}\left(\pi\right) & X\left(\frac{3}{4}\pi\right)\\
X^{2}\left(\frac{\pi}{4}\right) & z_{1}\left(\pi\right) & X^{2}\left(\frac{\pi}{4}\right) & M_{4}\ R_{4} & X^{2}\left(\frac{\pi}{4}\right)\\
z_{2}\left(\pi\right) & X^{2}\left(\frac{\pi}{4}\right) & Y\left(-\frac{3}{4}\pi\right) & z_{4}\left(\pi\right) & Y\left(\frac{\pi}{4}\right)\ M_{4}.
\end{array}\label{eq:Sequenz 3 Qubit BitFlip Messen}
\end{equation}
The graphical representation is shown in Fig.~\ref{fig:Three-qubit-bit-flip messen}.
For the optimization of this sequence, we actually used five qubits
in the computation (two auxiliary qubits). As a direct benefit, the
first measurement can be replaced by a swap gate which exchanges the
first auxiliary qubit with the second. In that way, both measurements
are postponed to the end of the sequence.

\begin{figure}
\includegraphics[width=0.845\columnwidth]{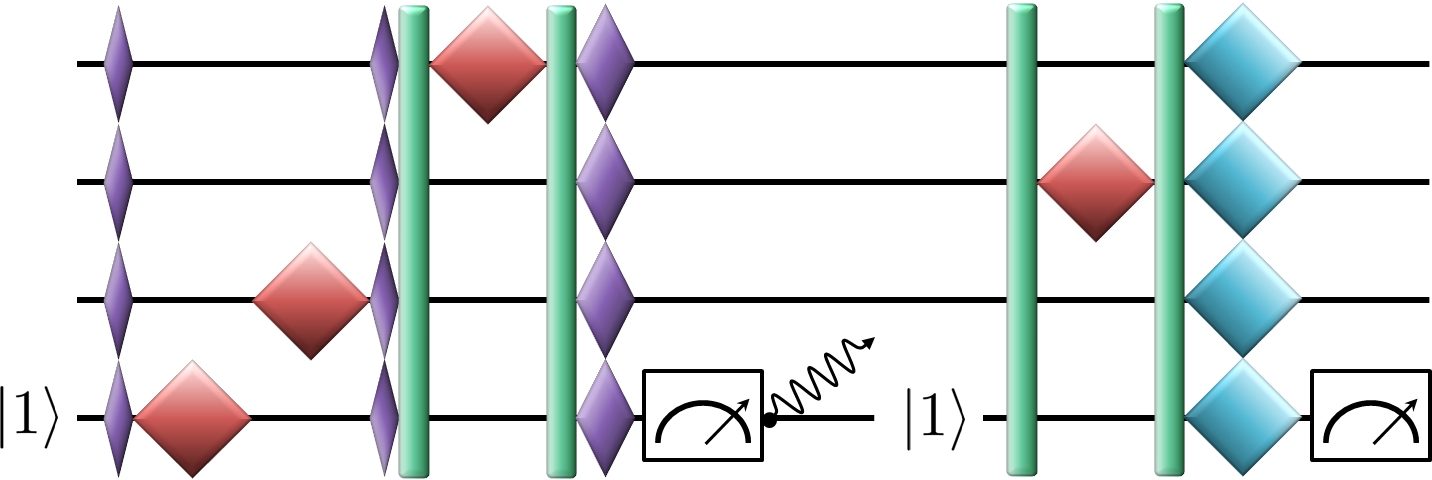}

\protect\caption{\label{fig:Three-qubit Phase messen}Measurement of the error syndrome
for the three qubit phase-flip error correction code \eqref{eq:Sequenz 3 Qubit Phase}.
The first three qubits encode the logical qubit. After the first measurement,
the auxiliary qubit is reset (wavy line).}
\end{figure}

For trapped ions, phase errors are usually a bigger issue than bit-flip
errors. Therefore, we also present an analog sequence to read out
the syndrome for phase errors

\begin{equation}
\begin{array}{lllll}
X\left(\frac{\pi}{4}\right) & z_{4}\left(\pi\right) & z_{3}\left(\pi\right) & X\left(-\frac{\pi}{4}\right) & X^{2}\left(\frac{\pi}{4}\right)\\
z_{1}\left(\pi\right) & X^{2}\left(\frac{\pi}{4}\right) & X\left(\frac{\pi}{2}\right) & M_{4}\ R_{4} & X^{2}\left(\frac{\pi}{4}\right)\\
z_{2}\left(\pi\right) & X^{2}\left(\frac{\pi}{4}\right) & Y\left(\pi\right) & M_{4},
\end{array}\label{eq:Sequenz 3 Qubit Phase}
\end{equation}
where the codewords for the logical qubit are $|0_{L}\rangle=|+,+,+\rangle$
and $|1_{L}\rangle=|-,-,-\rangle$. The graphical representation is
shown in Fig.~\ref{fig:Three-qubit Phase messen}.

\begin{figure}
\includegraphics[width=0.763\columnwidth]{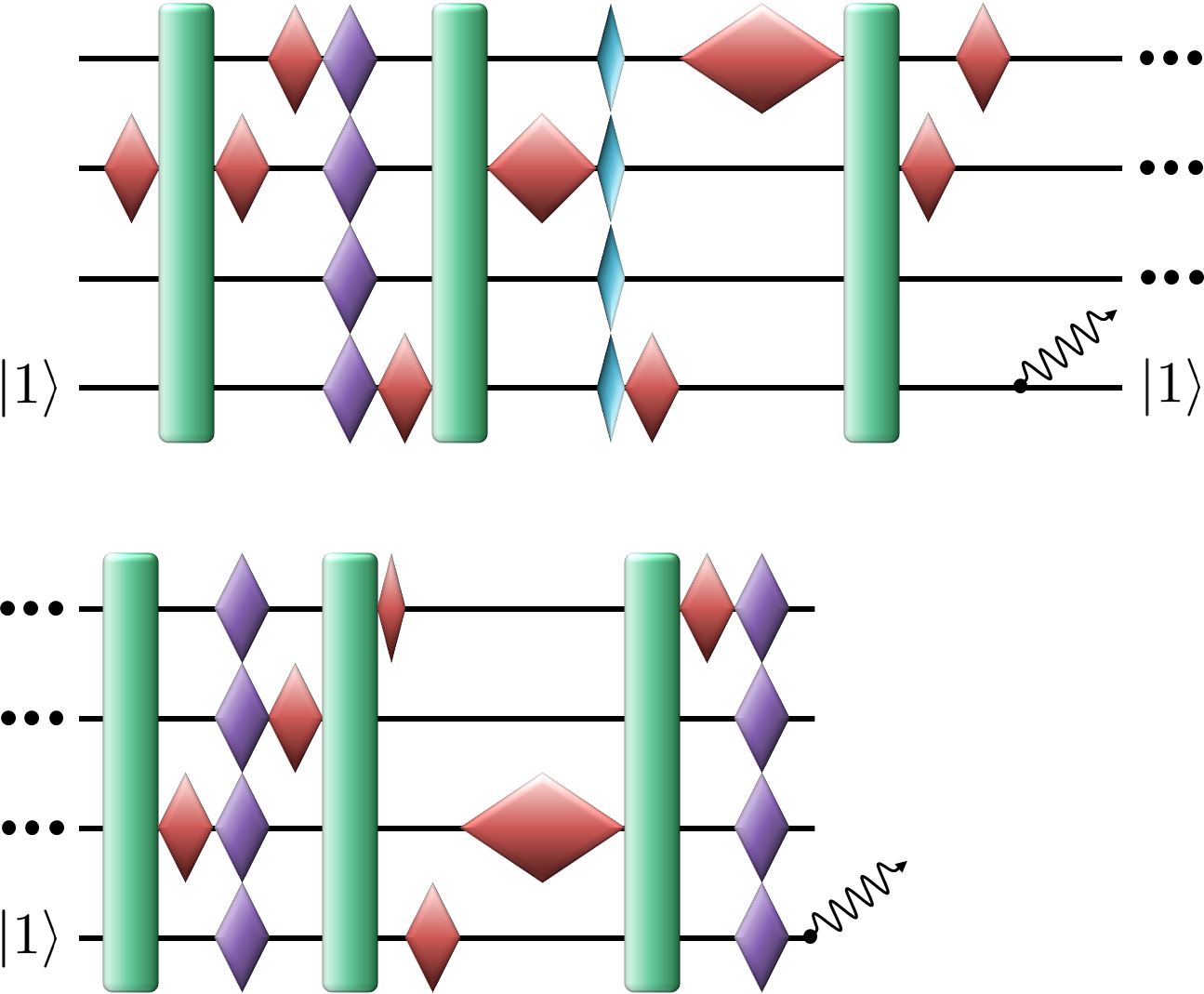}

\protect\caption{\label{fig:Three-qubit Fully Quantum}Coherent three qubit bit-flip
error correction code \eqref{eq:Sequenz 3 Qubit Fully Quantum}, which
corrects a possible error on the logical qubit (encoded in the first
three physical qubits) without the need of a measurement. To remove
the entropy, the auxiliary qubit has to be reset (wavy line). Because
of its length, the sequence is split into two parts. }
\end{figure}

Next, we look into a measurement-free coherent version of the bit-flip
error correction on four qubits. Fig.~\ref{fig:Three-qubit Fully Quantum}
shows suitable sequence for this task. Although no measurement is
needed in this case, the auxiliary qubit has to be reset to remove
the entropy from the qubit system (first reset is mandatory, second
is optional). For trapped ions, resetting a qubit can be done with
much less disturbance than a measurement. But this advantage comes
with a high price in the length of the sequence, 

\begin{equation}
\begin{array}{lllll}
z_{2}\left(\frac{\pi}{2}\right) & X^{2}\left(\frac{\pi}{2}\right) & z_{2}\left(\frac{\pi}{2}\right) & z_{1}\left(\frac{\pi}{2}\right) & X\left(\frac{\pi}{2}\right)\\
z_{4}\left(\frac{\pi}{2}\right) & X^{2}\left(\frac{\pi}{2}\right) & z_{2}\left(\pi\right) & Y\left(-\frac{\pi}{4}\right) & z_{4}\left(\frac{\pi}{2}\right)\\
z_{1}\left(-\frac{\pi}{2}\right) & X^{2}\left(\frac{\pi}{2}\right) & z_{2}\left(\frac{\pi}{2}\right) & z_{1}\left(\frac{\pi}{2}\right) & R_{4}\\
X^{2}\left(\frac{\pi}{2}\right) & z_{3}\left(\frac{\pi}{2}\right) & X\left(\frac{\pi}{2}\right) & z_{2}\left(\frac{\pi}{2}\right) & X^{2}\left(\frac{\pi}{2}\right)\\
z_{1}\left(\frac{\pi}{4}\right) & z_{4}\left(\frac{\pi}{2}\right) & z_{3}\left(-\frac{\pi}{2}\right) & X^{2}\left(\frac{\pi}{2}\right) & z_{1}\left(\frac{\pi}{2}\right)\\
X\left(\frac{\pi}{2}\right) & R_{4}.
\end{array}\label{eq:Sequenz 3 Qubit Fully Quantum}
\end{equation}
We mainly included this example to have a comparison between a measurement-based
and a measurement-free QEC code. Therefore, a similar structure was
adapted for both codes. For a practical realization, one might prefer
the measurement-free approach whose successful experimental realization
we presented in Ref.~\cite{Schindler2011}.

\subsection{Five qubit code and Steane code}

In this section, the five qubit code \cite{Laflamme1996} and the
Steane code \cite{Steane1996} are compared. Both are stabilizer codes
\cite{Nielsen2000,Gottesman1996} and promising candidates for future
experiments, since they allow us to detect arbitrary single qubit
errors. For both codes, we demonstrate how the logical qubit states
can be generated, how the stabilizers can be measured, and how some
nontrivial logical gates might be applied.

\subsubsection{State preparation for logical qubits}

The preparation of a predefined quantum state does not allow for any
variation except the global phase. Therefore, it is actually not the
best example to demonstrate the ideas outlined in Secs.~\ref{sec:Quantum-error-correction}
and \ref{sec:Optimization}. Still, since the preparation of logical
qubit is currently a hot topic and a necessary precondition for further
QEC operations, we think it is adequate to include some sequences
which can be used for this purpose.

\paragraph{For the five qubit code, }

the codeword for the logical zero state \cite{Nielsen2000} is
\begin{alignat}{2}
|0_{L}\rangle & = & \frac{1}{4}[ & |00000\rangle+|10010\rangle+|01001\rangle+|10100\rangle\nonumber \\
 &  & + & |01010\rangle-|11011\rangle-|00110\rangle-|11000\rangle\nonumber \\
 &  & - & |11101\rangle-|00011\rangle-|11110\rangle-|01111\rangle\nonumber \\
 &  & - & |10001\rangle-|01100\rangle-|10111\rangle+|00101\rangle].
\end{alignat}
The logical one state $|1_{L}\rangle$ is obtained from $|0_{L}\rangle$
by inverting all qubits ($0\leftrightarrow1).$ 

\begin{figure}
\includegraphics[width=0.68\columnwidth]{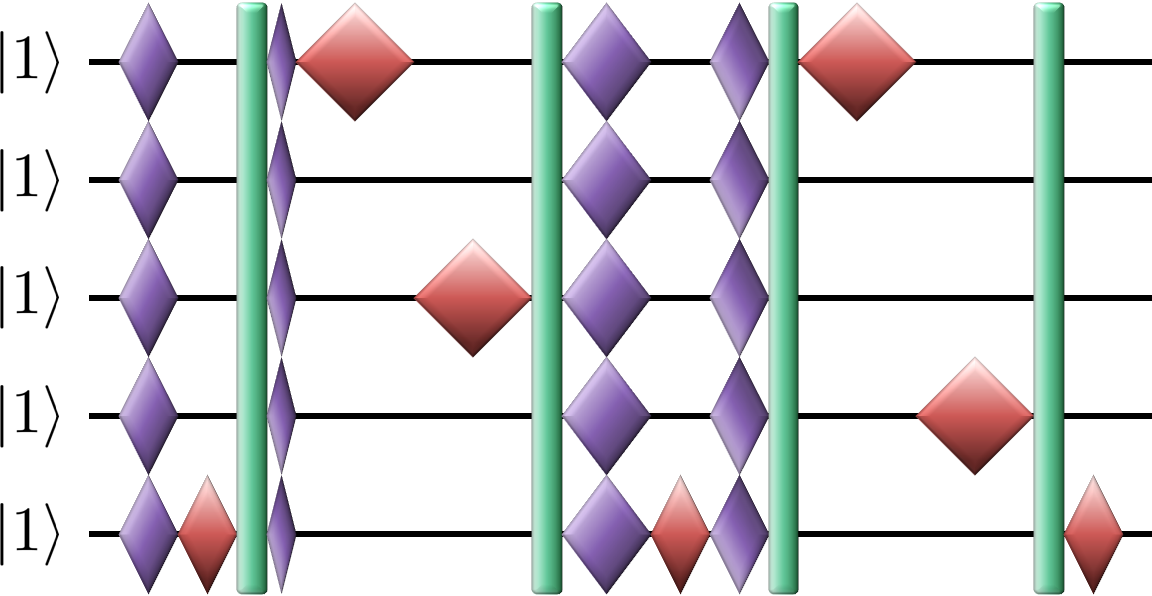}

\protect\caption{\label{fig:Sequence-to-generate 5Q mit 4 halben}Sequence to generate
the logical five qubit codeword $|0_{L}\rangle$ \eqref{eq:Sequence Zustand 5Q 4 halbe MS}.}

\end{figure}

Starting with the product state $|11111\rangle$, the logical zero
state $|0_{L}\rangle$ can be generated applying the following sequence
(Fig.~\ref{fig:Sequence-to-generate 5Q mit 4 halben}):

\begin{equation}
\begin{array}{lllll}
X\left(\frac{\pi}{2}\right) & z_{5}\left(\frac{\pi}{2}\right) & X^{2}\left(\frac{\pi}{4}\right) & X\left(-\frac{\pi}{4}\right) & z_{1}\left(\pi\right)\\
z_{3}\left(\pi\right) & X^{2}\left(\frac{\pi}{4}\right) & X\left(\frac{3}{4}\pi\right) & z_{5}\left(\frac{\pi}{2}\right) & X\left(-\frac{\pi}{2}\right)\\
X^{2}\left(\frac{\pi}{4}\right) & z_{1}\left(\pi\right) & z_{4}\left(\pi\right) & X^{2}\left(\frac{\pi}{4}\right) & z_{5}\left(\frac{\pi}{2}\right)
\end{array}\label{eq:Sequence Zustand 5Q 4 halbe MS}
\end{equation}
Replacing the first three $X(\varTheta)$ and the last $z_{5}\left(\frac{\pi}{2}\right)$
in this sequence by their inverse operations results in the logical
one state $|1_{L}\rangle$. 

\begin{figure}
\includegraphics[width=0.719\columnwidth]{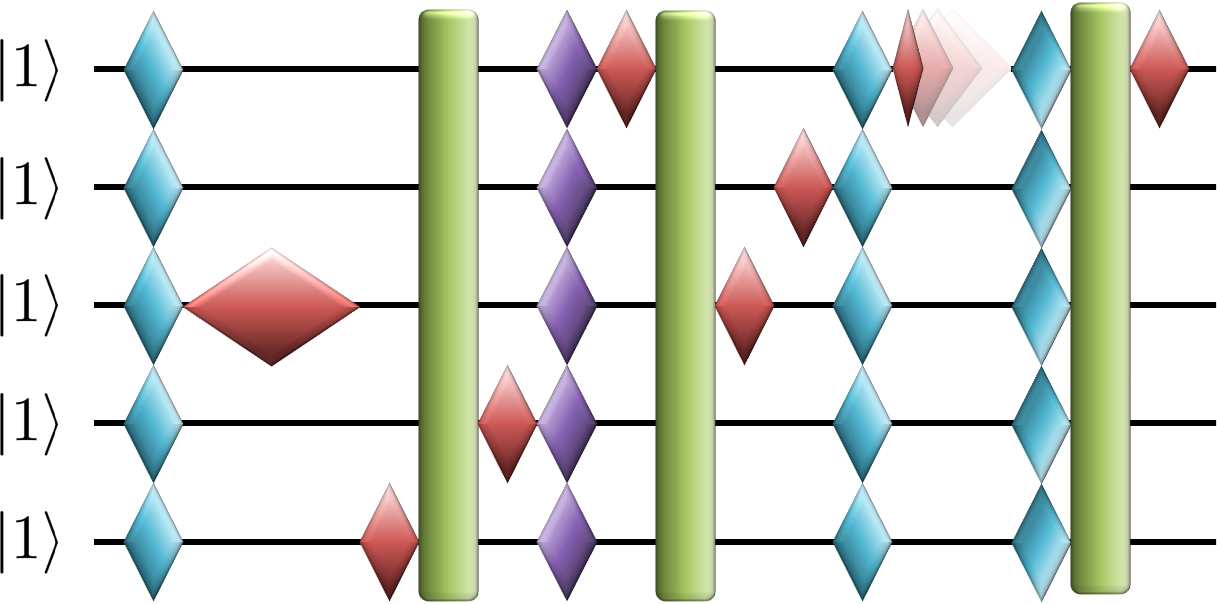}

\protect\caption{\label{fig:FiveQubitZustand}Sequence to generate the logical superposition
state $\sin\left(\alpha\right)\cdot|0_{L}\rangle+\cos\left(\alpha\right)\cdot|1_{L}\rangle$
for the five qubit code \eqref{eq:WinkelZustand 5 Qubit}. The length
$\varTheta$ of the second last $z_{1}(\varTheta)$ operation depends
on the angle $\alpha$, i.e., $z_{1}(\varTheta=2\alpha)$.}
\end{figure}

Alternatively, one can use the following sequence (Fig.~\ref{fig:FiveQubitZustand})
to obtain the logical superposition state $\sin\left(\alpha\right)\cdot|0_{L}\rangle+\cos\left(\alpha\right)\cdot|1_{L}\rangle$,
with arbitrary but known $\alpha$,

\begin{equation}
\begin{array}{lllll}
Y\left(\frac{\pi}{2}\right) & z_{3}\left(-\frac{\pi}{2}\right) & z_{5}\left(\frac{\pi}{2}\right) & Y^{2}\left(\frac{\pi}{2}\right) & z_{4}\left(\frac{\pi}{2}\right)\\
X\left(\frac{\pi}{2}\right) & z_{1}\left(\frac{\pi}{2}\right) & Y^{2}\left(\frac{\pi}{2}\right) & z_{3}\left(\frac{\pi}{2}\right) & z_{2}\left(\frac{\pi}{2}\right)\\
Y\left(\frac{\pi}{2}\right) & z_{1}\left(2\alpha\right) & Y\left(-\frac{\pi}{2}\right) & Y^{2}\left(\frac{\pi}{2}\right) & z_{1}\left(\frac{\pi}{2}\right)
\end{array}\label{eq:WinkelZustand 5 Qubit}
\end{equation}
If the first $Y\left(\frac{\pi}{2}\right)$ is replaced by $Y\left(-\frac{\pi}{2}\right)$
the result of the sequence is $\cos\left(\alpha\right)\cdot|0_{L}\rangle-\sin\left(\alpha\right)\cdot|1_{L}\rangle$.
This replacement is e.g.~of interest to produce the zero state $|0_{L}\rangle$,
which can now be done by setting $\alpha=0$. This allows us to omit
the subsequence $Y\left(\frac{\pi}{2}\right)\ z_{1}\left(2\alpha=0\right)\ Y\left(-\frac{\pi}{2}\right)$,
resulting in the shortest sequence found for $|0_{L}\rangle.$ 

The reader might wonder how the optimization for such a sequence with
an open parameter $\alpha$ was done. The trick is that such sequences
can be found using a \emph{fixed} $\alpha.$ To see this, we first
note that the optimal sequences which generate the states $|0_{L}\rangle$
respectively $|1_{L}\rangle$ are all built up by operations with
quantized arguments $\varTheta_{j}=\frac{m_{j}}{2^{n_{j}}}\cdot\pi$,
with $m_{j}\in\mathbb{Z}$, $n_{j}\in\mathbb{N}_{0}$. But if we choose
to generate the state $\sin\left(\alpha\right)\cdot|0_{L}\rangle+\cos\left(\alpha\right)\cdot|1_{L}\rangle$
with a fixed\emph{ }angle $\alpha$ that does not fit into this quantization
scheme (e.g.,\  $\alpha=\frac{\pi}{5}$), at least one operation
is needed with an argument $\varTheta_{k}\neq\frac{m}{2^{n}}\cdot\pi$.
Looking into various different sequences that contain exactly one
such operation with an argument $\varTheta_{k}\neq\frac{m}{2^{n}}\cdot\pi$,
we found that for some of these sequences the argument $\varTheta_{k}=\varTheta_{k}(\alpha)$
could be expressed as a  (linear) function of $\alpha$. In other
words, some sequences, which were found for a specific value of $\alpha$,
can be generalized to arbitrary values of $\alpha$ by adopting the
argument $\varTheta_{k}(\alpha)$, while all other arguments $\varTheta_{j\neq k}$
stay unchanged, as in Eq.~\eqref{eq:WinkelZustand 5 Qubit} with
its $z_{1}\left(\varTheta=2\alpha\right)$ operation.

If we try to introduce a further phase factor $\sin\left(\alpha\right)\cdot|0_{L}\rangle+e^{i\beta}\cos\left(\alpha\right)\cdot|1_{L}\rangle$
we need at least two operations with arguments $\varTheta_{k}\neq\frac{m}{2^{n}}\cdot\pi$.
We found several of such sequences which are generalizable to arbitrary
$\alpha$ and $\beta$, but the values of the two corresponding arguments
$\varTheta_{k}(\alpha,\beta)$ have to be determined numerically.
None of them shows a simple linear relation $\varTheta_{k}(\alpha,\beta)\propto\alpha,\beta$
nor any other evident analytical dependence on the chosen angles $\alpha,\beta$,
which would allow us to present these results.

\begin{figure}
\includegraphics[width=0.984\columnwidth]{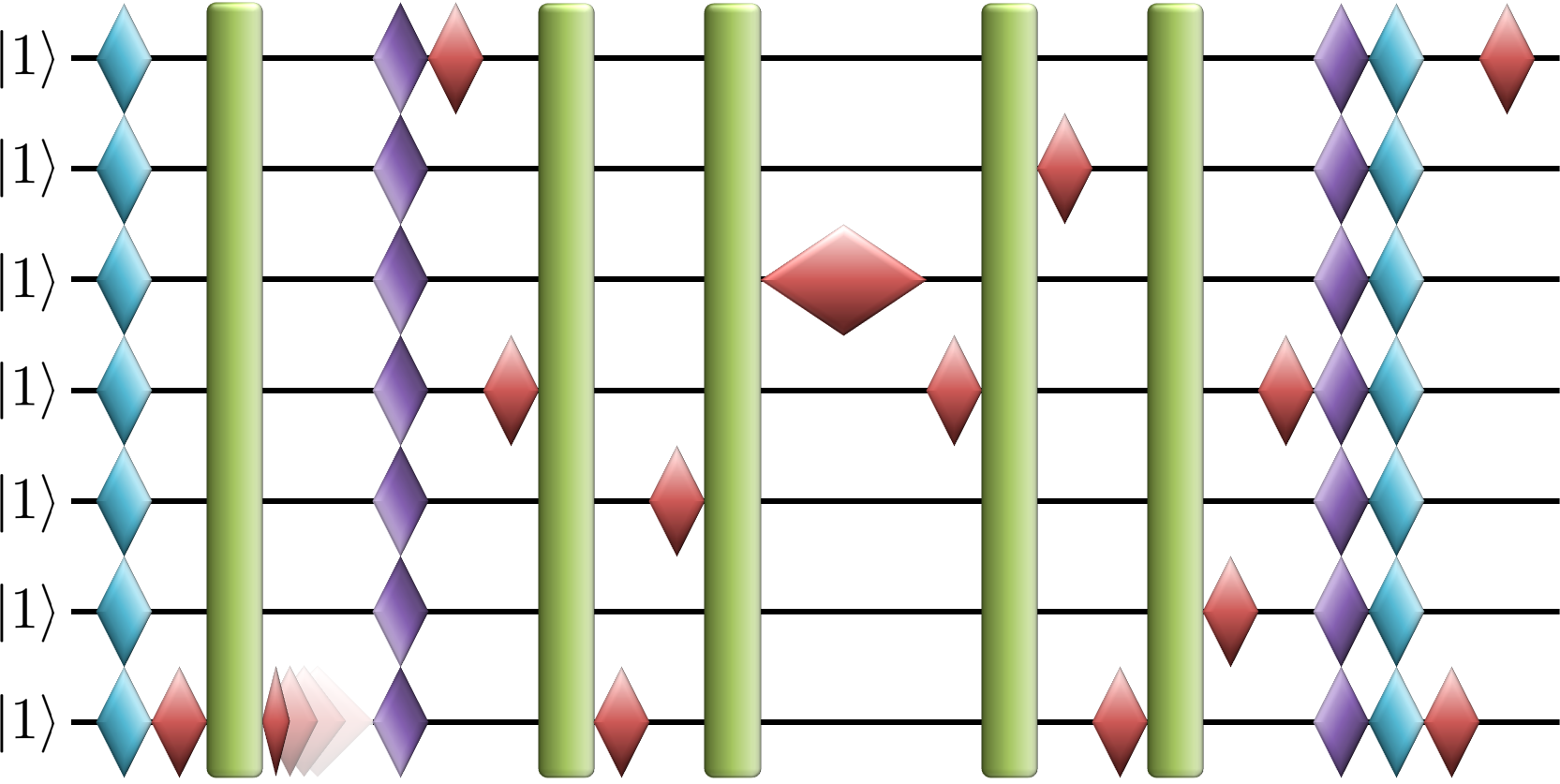}

\protect\caption{\label{fig:Sequence-to-generate Steane Winkel Zustand}Sequence to
generate the logical superposition state $\cos\left(\alpha\right)\cdot|0_{L}\rangle+\sin\left(\alpha\right)\cdot|1_{L}\rangle$
for the Steane code \eqref{eq:Sequenz Stean Winkel}. The length $\varTheta$
of the second $z_{7}(\varTheta)$ operation depends on the angle $\alpha$,
i.e.,\ $z_{7}(\varTheta=2\alpha-\frac{\pi}{2})$.}
\end{figure}

\paragraph{For the Steane code,}

the codeword for the logical zero state \cite{Nielsen2000} is 
\begin{alignat}{2}
|0_{L}\rangle & = & \frac{1}{\sqrt{8}}\big[ & |0000000\rangle+|1010101\rangle+|0110011\rangle\nonumber \\
 &  & + & |1100110\rangle+|0001111\rangle+|1011010\rangle\nonumber \\
 &  & + & |0111100\rangle+|1101001\rangle\big]
\end{alignat}
and the logical one state $|1_{L}\rangle$ can be obtained from $|0_{L}\rangle$
by inverting all qubits ($0\leftrightarrow1).$ Analog to the five
qubit code, we can list a sequence (Fig.~\ref{fig:Sequence-to-generate Steane Winkel Zustand})
to produce the logical superposition state $\cos\left(\alpha\right)\cdot|0_{L}\rangle+\sin\left(\alpha\right)\cdot|1_{L}\rangle$
out of the product state $|1111111\rangle$,
\begin{figure}
\includegraphics[width=0.917\columnwidth]{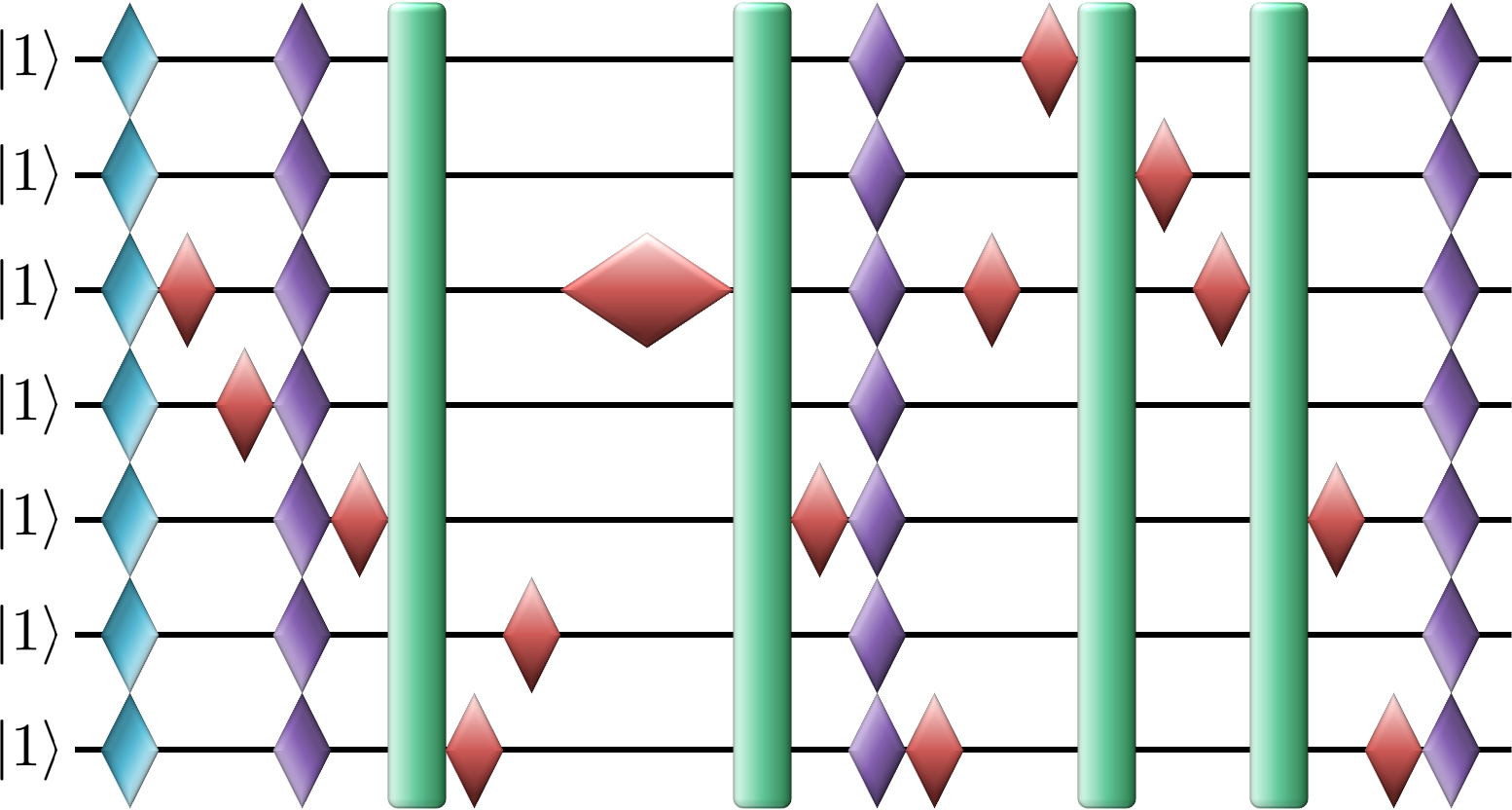}\protect\caption{\label{fig:Sequence-Zustand Steane}Sequence to generate the logical
Steane codeword $|0_{L}\rangle$ \eqref{eq:Sequence Zustand Steane}. }
\end{figure}

\begin{equation}
\begin{array}{lllll}
Y\left(\frac{\pi}{2}\right) & z_{7}\left(\frac{\pi}{2}\right) & Y^{2}\left(\frac{\pi}{2}\right) & z_{7}\left(2\alpha-\frac{\pi}{2}\right) & X\left(-\frac{\pi}{2}\right)\\
z_{1}\left(\frac{\pi}{2}\right) & z_{4}\left(\frac{\pi}{2}\right) & Y^{2}\left(\frac{\pi}{2}\right) & z_{7}\left(\frac{\pi}{2}\right) & z_{5}\left(\frac{\pi}{2}\right)\\
Y^{2}\left(\frac{\pi}{2}\right) & z_{3}\left(-\frac{\pi}{2}\right) & z_{4}\left(\frac{\pi}{2}\right) & Y^{2}\left(\frac{\pi}{2}\right) & z_{2}\left(\frac{\pi}{2}\right)\\
z_{7}\left(\frac{\pi}{2}\right) & Y^{2}\left(\frac{\pi}{2}\right) & z_{6}\left(\frac{\pi}{2}\right) & z_{4}\left(\frac{\pi}{2}\right) & X\left(\frac{\pi}{2}\right)\\
Y\left(\frac{\pi}{2}\right) & z_{7}\left(\frac{\pi}{2}\right) & z_{1}\left(\frac{\pi}{2}\right).
\end{array}\label{eq:Sequenz Stean Winkel}
\end{equation}
Contrary to the five qubit code, where the shortest sequences for
$|0_{L}\rangle$ and $|1_{L}\rangle$ could be found by setting $\alpha=0$
in Eq.~\eqref{eq:WinkelZustand 5 Qubit}, for the Steane code it
is better to resort to alternative sequences for the logical codewords
$|0_{L}\rangle$ and $|1_{L}\rangle$. We found sequences consisting
of only 19 operations, but with five Mølmer-Sørensen (MS) gates ($X^{2}\left(\frac{\pi}{2}\right)$
or $Y^{2}\left(\frac{\pi}{2}\right)$), as above \eqref{eq:Sequenz Stean Winkel}.
Since the MS gates are the main source for losses in fidelity, another
sequence to create the logical zero state $|0_{L}\rangle$ with 22
operations but only four MS gates might be more favorable (Fig.~\ref{fig:Sequence-Zustand Steane}),

\begin{equation}
\begin{array}{lllll}
Y\left(-\frac{\pi}{2}\right) & z_{3}\left(\frac{\pi}{2}\right) & z_{4}\left(\frac{\pi}{2}\right) & X\left(-\frac{\pi}{2}\right) & z_{5}\left(\frac{\pi}{2}\right)\\
X^{2}\left(\frac{\pi}{2}\right) & z_{7}\left(\frac{\pi}{2}\right) & z_{6}\left(\frac{\pi}{2}\right) & z_{3}\left(-\frac{\pi}{2}\right) & X^{2}\left(\frac{\pi}{2}\right)\\
z_{5}\left(\frac{\pi}{2}\right) & X\left(\frac{\pi}{2}\right) & z_{7}\left(\frac{\pi}{2}\right) & z_{3}\left(\frac{\pi}{2}\right) & z_{1}\left(\frac{\pi}{2}\right)\\
X^{2}\left(\frac{\pi}{2}\right) & z_{2}\left(\frac{\pi}{2}\right) & z_{3}\left(\frac{\pi}{2}\right) & X^{2}\left(\frac{\pi}{2}\right) & z_{5}\left(\frac{\pi}{2}\right)\\
z_{7}\left(\frac{\pi}{2}\right) & X\left(-\frac{\pi}{2}\right).
\end{array}\label{eq:Sequence Zustand Steane}
\end{equation}
Replacing the first operator $Y\left(-\frac{\pi}{2}\right)$ by its
inverse $Y\left(\frac{\pi}{2}\right)$ allows us to generate the logical
one state $|1_{L}\rangle$.

\begin{figure}
\includegraphics[width=0.862\columnwidth]{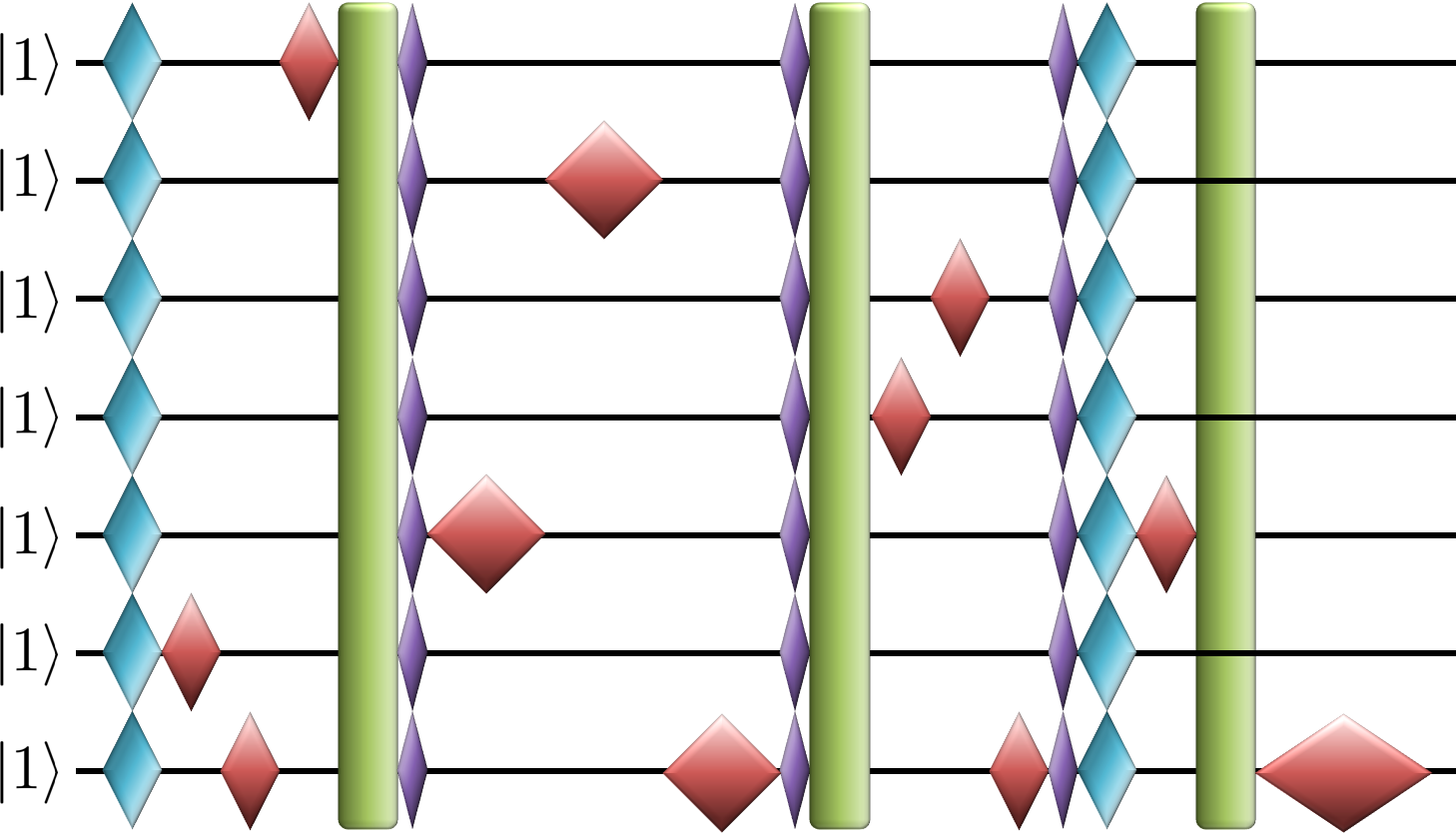}\protect\caption{\label{fig:Steane Null 3MS special}Sequence to generate the logical
Steane codeword $|0_{L}\rangle$ \eqref{eq:Steane 3MS special}. The
third Mølmer-Sørensen gate only acts on the qubits number 1, 3, 5,
and 7.}
\end{figure}

In Ref.~\cite{Nigg2014}, a logical state (color-code) is created
with the help of MS gates which operate solely on subsets of the entire
seven qubits. Inspired by this work, we present a further alternative
to create the $|0_{L}\rangle$ state. Here, we come along with three
MS gates $Y^{2}\left(\frac{\pi}{2}\right)$, where the last (third)
MS gate $\tilde{Y}_{1,3,5,7}^{2}\left(\frac{\pi}{2}\right)$ operates
only on the qubits number 1, 3, 5, and 7 (Fig.~\ref{fig:Steane Null 3MS special}),

\begin{equation}
\begin{array}{lllll}
Y\left(-\frac{\pi}{2}\right) & z_{6}\left(\frac{\pi}{2}\right) & z_{7}\left(\frac{\pi}{2}\right) & z_{1}\left(\frac{\pi}{2}\right) & Y^{2}\left(\frac{\pi}{2}\right)\\
X\left(\frac{\pi}{4}\right) & z_{5}\left(\pi\right) & z_{2}\left(\pi\right) & z_{7}\left(\pi\right) & X\left(\frac{\pi}{4}\right)\\
Y^{2}\left(\frac{\pi}{2}\right) & z_{4}\left(\frac{\pi}{2}\right) & z_{3}\left(\frac{\pi}{2}\right) & z_{7}\left(\frac{\pi}{2}\right) & X\left(\frac{\pi}{4}\right)\\
Y\left(-\frac{\pi}{2}\right) & z_{5}\left(\frac{\pi}{2}\right) & \tilde{Y}_{1,3,5,7}^{2}\left(\frac{\pi}{2}\right) & z_{7}\left(-\frac{\pi}{2}\right).
\end{array}\label{eq:Steane 3MS special}
\end{equation}
Some more comments concerning MS gates on subsets can be found in
App.~\ref{sec:Elementary-operations-for-Trapped-ions}.

\subsubsection{Stabilizer\label{sub:Stabilizer}}

The five qubit code and the Steane code are examples for stabilizer
codes. A stabilizer is a product of local Pauli operators and the
identity, e.g.,~the first (of four) stabilizer of the five qubit
code is given by $XZZXI$ (which is shorthand for $\sigma_{x}^{(1)}\cdot\sigma_{z}^{(2)}\cdot\sigma_{z}^{(3)}\cdot\sigma_{x}^{(4)}\cdot{\mathbbm1}^{(5)})$;
see Tab.~\ref{tab:Five Qubit Stabilizer}. The measurement result
of the stabilizers does not reveal the state of the logical qubit,
but it allows us to infer which error might have happened. To measure
a stabilizer, its result is mapped onto an auxiliary qubit, which
then can be read out. In short: the situation complies with the measurement
of the syndrome discussed in Sec.~\ref{sub:Measuring-the-error syndrom}.

\begin{table}
\begin{tabular}{c|c}
Number & Operator\tabularnewline
\hline 
1 & $XZZXI$\tabularnewline
2 & $IXZZX$\tabularnewline
3 & $XIXZZ$\tabularnewline
4 & $ZXIXZ$\tabularnewline
\end{tabular}\protect\caption{\label{tab:Five Qubit Stabilizer}Description of the four stabilizers
for the five qubit code.}
\end{table}

\begin{table}
\begin{tabular}{c|c}
Number & Operator\tabularnewline
\hline 
1 & $IIIXXXX$\tabularnewline
2 & $IXXIIXX$\tabularnewline
3 & $XIXIXIX$\tabularnewline
4 & $IIIZZZZ$\tabularnewline
5 & $IZZIIZZ$\tabularnewline
6 & $ZIZIZIZ$\tabularnewline
\end{tabular}\protect\caption{\label{tab:Steane Stabilizer}Description of the six stabilizers for
the Steane code.}
\end{table}

\begin{figure}

\includegraphics[width=0.628\columnwidth]{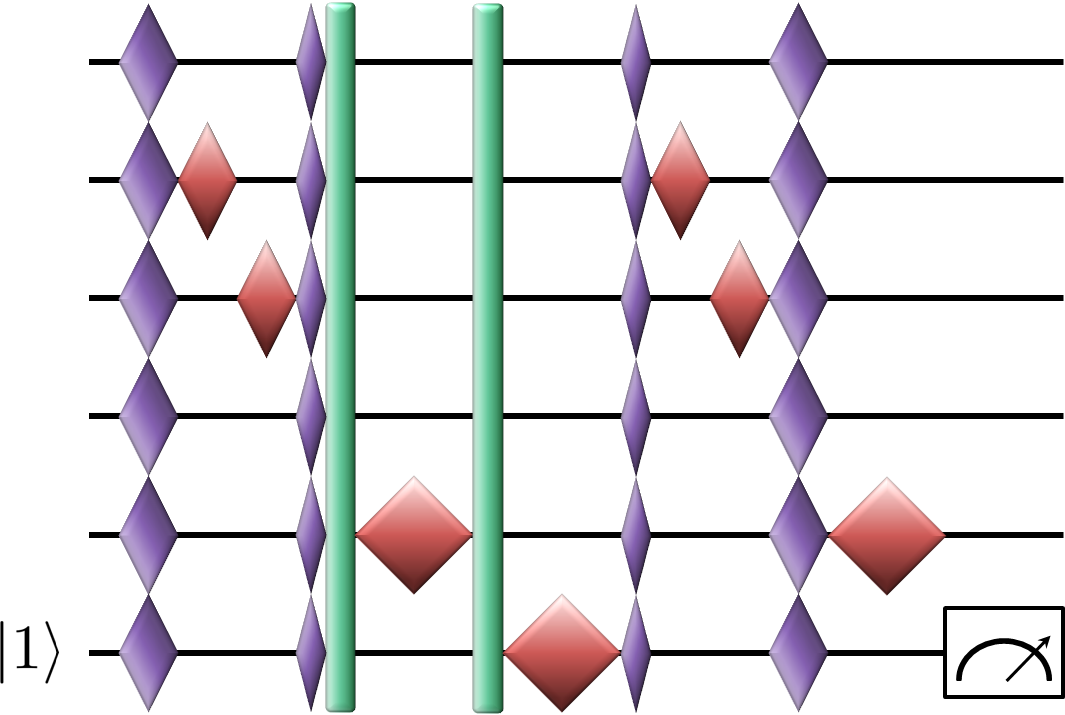}\protect\caption{\label{fig:Stabilzer-Five Qubit}Measurement of the stabilizer ($XZZXI$)
for the five qubit code \eqref{eq:Stabilizer Five Qubit}.}

\end{figure}

\paragraph{For the five qubit code,}

the four stabilizers (Tab.~\ref{tab:Five Qubit Stabilizer}) are
related by cyclical permutation. Therefore, it suffices to list the
sequence (Fig.~\ref{fig:Stabilzer-Five Qubit}) of the first stabilizer
($XZZXI$) \--- the others follow by cyclical permutation of the
local $z_{j}(\varTheta)$ operators in the sequence (not including
the operator $z_{6}\left(\pi\right)$ acting on the auxiliary qubit!)
\begin{equation}
\begin{array}{lllll}
X\left(-\frac{\pi}{2}\right) & z_{2}\left(\frac{\pi}{2}\right) & z_{3}\left(\frac{\pi}{2}\right) & X\left(\frac{\pi}{4}\right) & X^{2}\left(\frac{\pi}{4}\right)\\
z_{5}\left(\pi\right) & X^{2}\left(\frac{\pi}{4}\right) & z_{6}\left(\pi\right) & X\left(\frac{\pi}{4}\right) & z_{2}\left(\frac{\pi}{2}\right)\\
z_{3}\left(\frac{\pi}{2}\right) & X\left(-\frac{\pi}{2}\right) & z_{5}\left(\pi\right) & M_{6}.
\end{array}\label{eq:Stabilizer Five Qubit}
\end{equation}

\begin{figure}
\includegraphics[width=0.949\columnwidth]{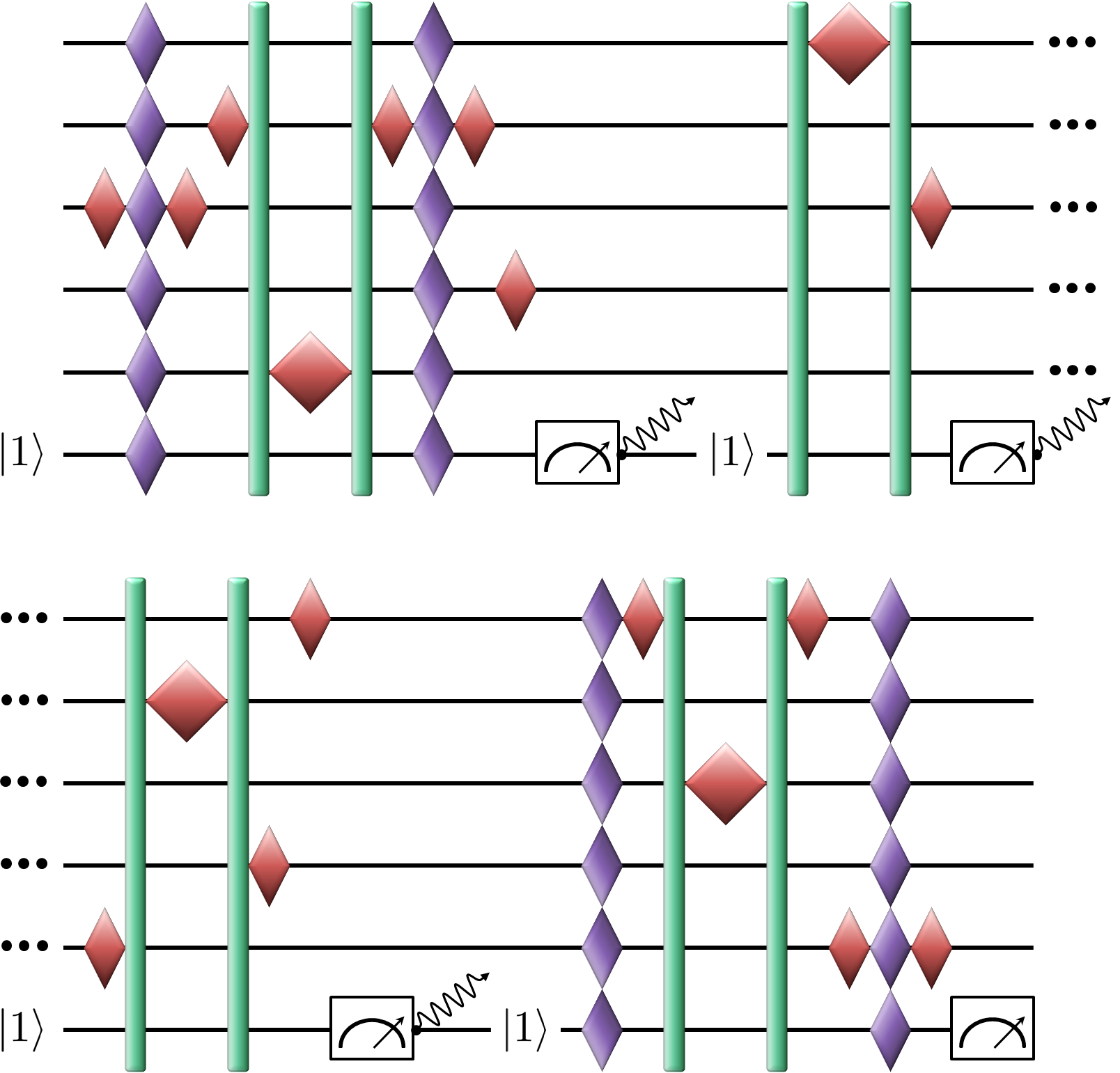}

\protect\caption{\label{fig:Measurement-of-all foir Stabilizer}Measurement of all
four stabilizers for the five qubit code \eqref{eq:All_Stabilizer}.
Because of its length, the sequence is split into two parts. After
the first three measurements, the auxiliary qubit is reset.}

\end{figure}

To detect an error, all four stabilizers have to be measured. Instead
of applying all four stabilizer sequences individually, one might
also use the following sequence to get the four results (Fig.~\ref{fig:Measurement-of-all foir Stabilizer}):

\begin{equation}
\begin{array}{lllll}
z_{3}\left(\frac{\pi}{2}\right) & X\left(\frac{\pi}{2}\right) & z_{3}\left(\frac{\pi}{2}\right) & z_{2}\left(\frac{\pi}{2}\right) & X^{2}\left(\frac{\pi}{4}\right)\\
z_{5}\left(\pi\right) & X^{2}\left(\frac{\pi}{4}\right) & z_{2}\left(\frac{\pi}{2}\right) & X\left(-\frac{\pi}{2}\right) & z_{2}\left(\frac{\pi}{2}\right)\\
z_{4}\left(\frac{\pi}{2}\right) & M_{6}\ R_{6} & X^{2}\left(\frac{\pi}{4}\right) & z_{1}\left(\pi\right) & X^{2}\left(\frac{\pi}{4}\right)\\
z_{3}\left(\frac{\pi}{2}\right) & M_{6}\ R_{6} & z_{5}\left(\frac{\pi}{2}\right) & X^{2}\left(\frac{\pi}{4}\right) & z_{2}\left(\pi\right)\\
X^{2}\left(\frac{\pi}{4}\right) & z_{4}\left(\frac{\pi}{2}\right) & z_{1}\left(\frac{\pi}{2}\right) & M_{6}\ R_{6} & X\left(-\frac{\pi}{2}\right)\\
z_{1}\left(\frac{\pi}{2}\right) & X^{2}\left(\frac{\pi}{4}\right) & z_{3}\left(\pi\right) & X^{2}\left(\frac{\pi}{4}\right) & z_{1}\left(\frac{\pi}{2}\right)\\
z_{5}\left(\frac{\pi}{2}\right) & X\left(\frac{\pi}{2}\right) & z_{5}\left(\frac{\pi}{2}\right) & M_{6}.
\end{array}\label{eq:All_Stabilizer}
\end{equation}

Here, we actually used nine qubits for the optimization and replaced
the first three measurements by swap gates to avoid measurements within
the sequence. This is basically the same trick as described for Eq.~\eqref{eq:Sequenz 3 Qubit BitFlip Messen}. 

The sequence we found consists of only $30$ unitary operations, which
has to be compared with $4\times13=52$ for the sum of the unitary
operations of all four individual stabilizers. The ratio of the needed
operations becomes even more impressive when we take into account
that the number of the Mølmer-Sørensen (MS) gates $X^{2}(\frac{\pi}{4})$
could not be altered. To entangle two formerly unentangled qubits,
one needs at least two $X^{2}(\frac{\pi}{4})$ gates. Since the auxiliary
qubit has to be entangled four times with the code qubits, the minimal
number of $X^{2}(\frac{\pi}{4})$ gates is eight. With that in mind,
the comparison to be made is $30-8=22$ versus $52-8=44$, which is
a factor of two! 

Contrary to all other sequences presented in this paper, this is the
only one which is not the result of an ab initio calculation starting
from a random sequence. The start sequence was a composite of previous
results for the single stabilizers, which were optimized with the
help of the method described in App.~\prettyref{sub:Disturbing-the-sequence}. 

With an increasing complexity of future quantum algorithms we might
face much more tasks which can only be solved with the help of composite
sequences. Therefore, it seems very promising that we managed to reduce
the number of the non Mølmer-Sørensen gates by a factor of two. Unfortunately,
a great part of the savings might only be due to the gauge freedom
and the fact that the four stabilizers are realized by quite similar
quantum circuits. If this is the case, the four-stabilizer-sequence
is just a positive exception. Anyway, it still seems a realistic hope
that similar savings could be achievable when the four stabilizers
are optimized for other quantum systems.

\begin{figure}

\includegraphics[width=0.836\columnwidth]{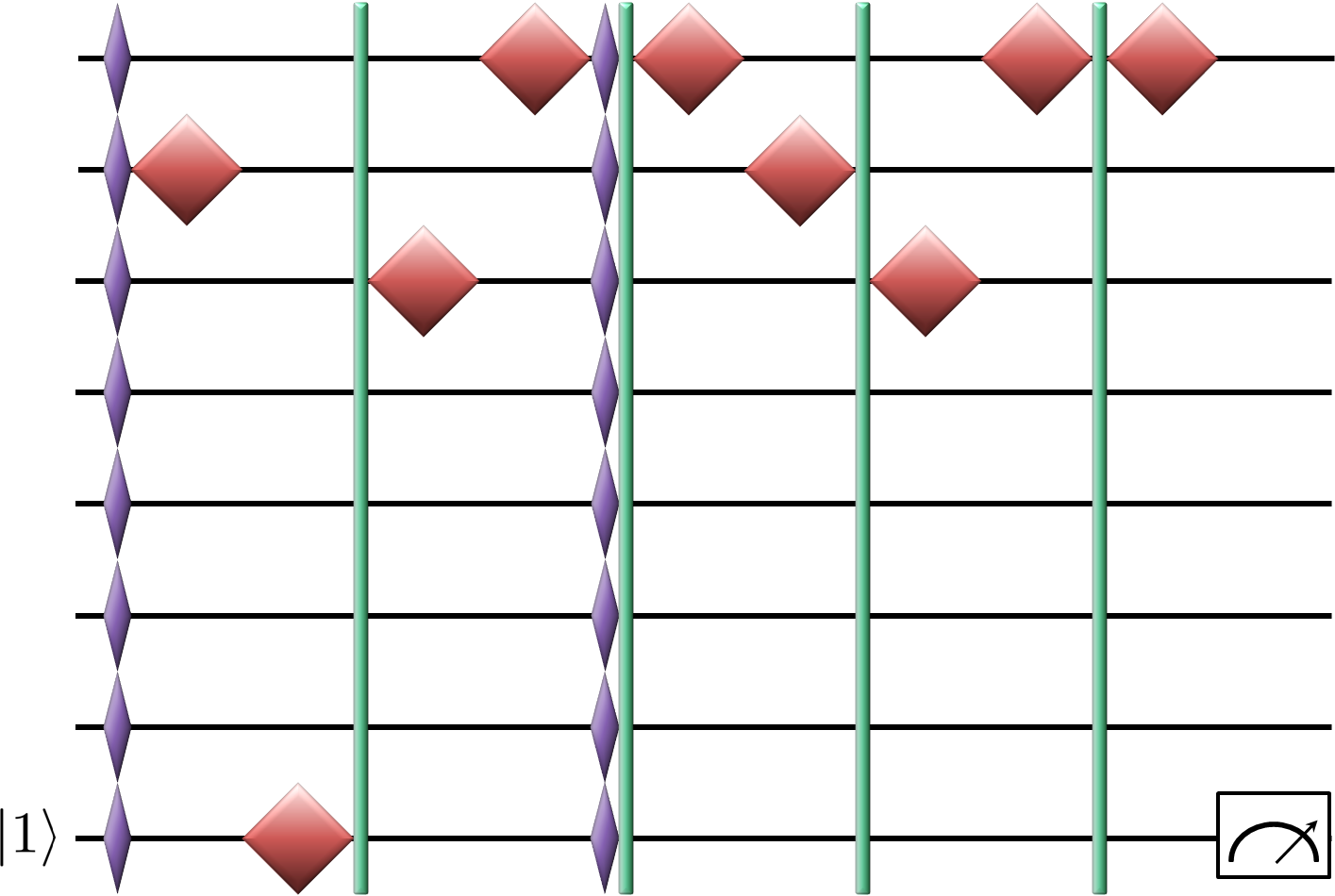}\protect\caption{\label{fig:Sequenz Stabilizer 1 Steane}Measurement of the first stabilizer
($IIIXXXX$) for the Steane code \eqref{eq:Sequence Stabilizer Steane 1}. }
\end{figure}

\paragraph{For the Steane code, }

the sequences for all six stabilizers (Tab.~\ref{tab:Steane Stabilizer})
can be derived from two base sequences. To measure the first stabilizer
($IIIXXXX$), one can use the sequence (Fig.~\ref{fig:Sequenz Stabilizer 1 Steane}):
\begin{equation}
\begin{array}{lllll}
X\left(\frac{\pi}{4}\right) & z_{2}\left(\pi\right) & z_{8}\left(\pi\right) & X^{2}\left(\frac{\pi}{8}\right) & z_{3}\left(\pi\right)\\
z_{1}\left(\pi\right) & X\left(\frac{\pi}{4}\right) & X^{2}\left(\frac{\pi}{8}\right) & z_{1}\left(\pi\right) & z_{2}\left(\pi\right)\\
X^{2}\left(\frac{\pi}{8}\right) & z_{3}\left(\pi\right) & z_{1}\left(\pi\right) & X^{2}\left(\frac{\pi}{8}\right) & z_{1}\left(\pi\right)\ M_{8}.
\end{array}\label{eq:Sequence Stabilizer Steane 1}
\end{equation}
Observe that the local $z_{j}(\pi)$ operators act only on positions
where the stabilizer has an identity $I$ (and on the auxiliary qubit).
To get the second and the third stabilizer (see Tab.~\ref{tab:Steane Stabilizer}),
the only thing to do is to rearrange the $z_{j}(\pi)$ accordingly
to the identities in these stabilizers 
\begin{eqnarray}
\underbrace{\left(z_{1}(\pi);z_{2}(\pi);z_{3}(\pi)\right)}_{\textrm{first stabilizer}} & \rightarrow & \underbrace{\left(z_{1}(\pi);z_{4}(\pi);z_{5}(\pi)\right)}_{\textrm{second stabilizer}}\nonumber \\
 & \rightarrow & \underbrace{\left(z_{2}(\pi);z_{4}(\pi);z_{6}(\pi)\right)}_{\textrm{third stabilizer}}.
\end{eqnarray}
\begin{figure}
\includegraphics[width=0.871\columnwidth]{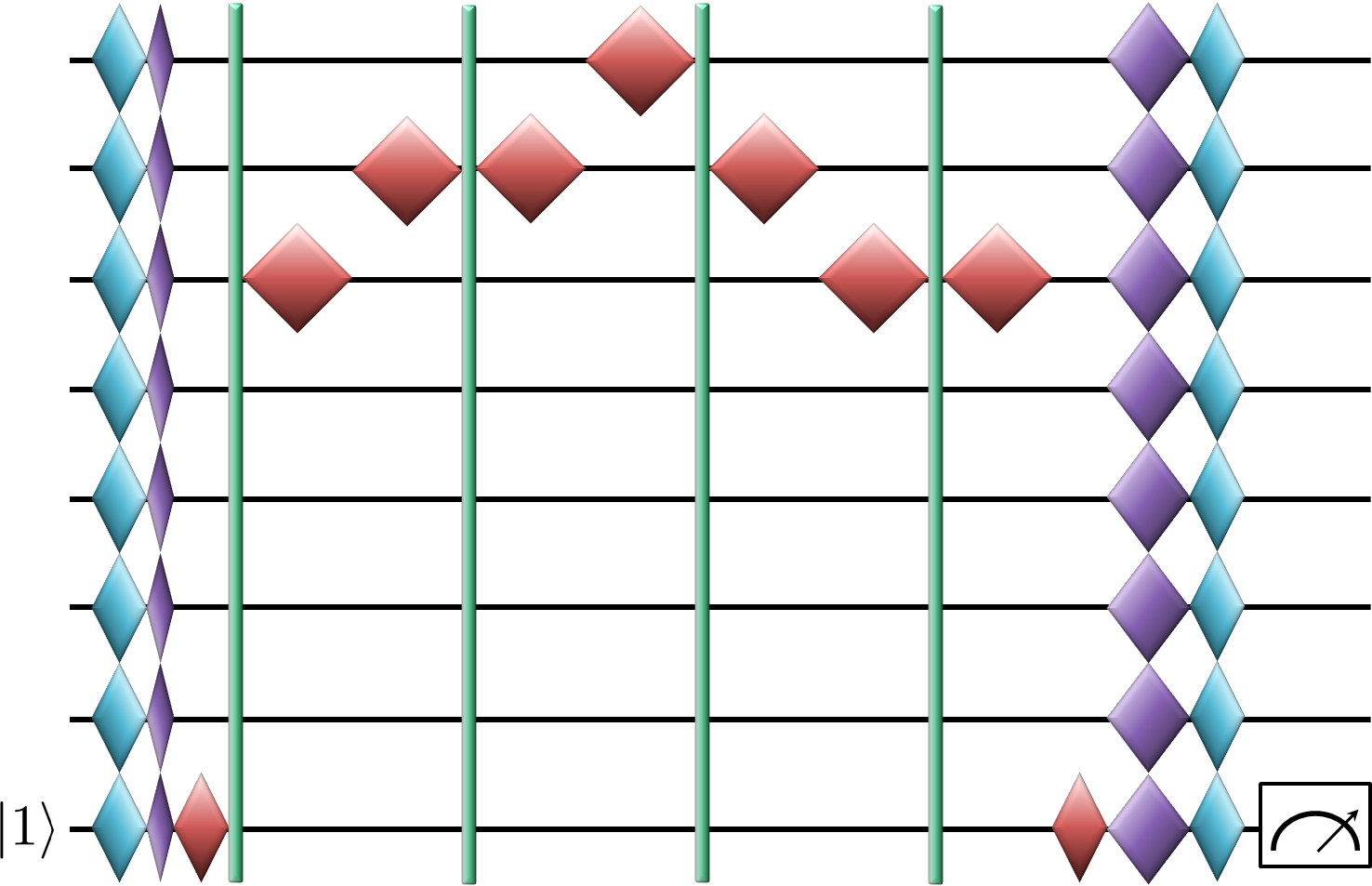}

\protect\caption{\label{fig:Sequenz Stabilizer 4 Steane}Measurement of the fourth
stabilizer ($IIIZZZZ$) for the Steane code \eqref{eq:Sequenz Steane Stabilizer 4}.}
\end{figure}
This rearrangement might not be trivial, but still, it works. Moreover,
the same rearrangement trick of the $z_{j}(\pi)$ can also be applied
to derive the sequences for the fifth and sixth stabilizer from the
sequence of the fourth stabilizer, which is given by (Fig.~\ref{fig:Sequenz Stabilizer 4 Steane})
\begin{equation}
\begin{array}{lllll}
Y\left(\frac{\pi}{2}\right) & X\left(-\frac{\pi}{4}\right) & z_{8}\left(\frac{\pi}{2}\right) & X^{2}\left(\frac{\pi}{8}\right) & z_{3}\left(\pi\right)\\
z_{2}\left(\pi\right) & X^{2}\left(\frac{\pi}{8}\right) & z_{2}\left(\pi\right) & z_{1}\left(\pi\right) & X^{2}\left(\frac{\pi}{8}\right)\\
z_{2}\left(\pi\right) & z_{3}\left(\pi\right) & X^{2}\left(\frac{\pi}{8}\right) & z_{3}\left(\pi\right) & z_{8}\left(\frac{\pi}{2}\right)\\
X\left(\frac{3}{4}\pi\right) & Y\left(\frac{\pi}{2}\right) & M_{8}.
\end{array}\label{eq:Sequenz Steane Stabilizer 4}
\end{equation}

It goes without saying that alternative sequences can be found, as
well. Here, we presented solutions for the five qubit code and the
Steane code which contain the minimal total length $\sum_{X^{2}}|\varTheta|$
of Mølmer-Sørensen (MS) gates $X^{2}(\varTheta)$ needed to entangle
the formerly unentangled auxiliary qubit with the code qubits: The
Steane code needs four $X^{2}\left(\varTheta=\frac{\pi}{8}\right)$
gates and the five qubit code two $X^{2}\left(\varTheta=\frac{\pi}{4}\right)$
gates, which in both cases sums up to the minimal value $\sum_{X^{2}}|\varTheta|=\frac{\pi}{2}$.
For $\sum_{X^{2}}|\varTheta|<\frac{\pi}{2}$, only partial entanglement
can be achieved, while $\sum_{X^{2}}|\varTheta|>\frac{\pi}{2}$ is
possible as well.

For a practical realization on the other hand, other aspects than
the length $\sum_{X^{2}}|\varTheta|$ might become more important.
In experiments, one might want to apply a stabilizer sequence after
a state preparation sequence. The state preparation sequences for
the Steane code \eqref{eq:Sequenz Stean Winkel} \eqref{eq:Sequence Zustand Steane}
are built on $Y^{2}(\frac{\pi}{2})$ and $X^{2}(\frac{\pi}{2})$ gates,
while the stabilizer sequence uses $X^{2}\left(\frac{\pi}{8}\right)$
gates. As long as it is experimentally advisable to use just one type
of MS gate, these two sequences are not a perfect match. Under this
condition, it seems favorable to resort to other sequences, which
might appear suboptimal when treated as isolated. Alternatively, one
could also consider to replace each $X^{2}\left(\frac{\pi}{2}\right)$
by four consecutive $X^{2}\left(\frac{\pi}{8}\right)$. Usually, such
a replacement also entails a certain loss in the experimental fidelity.

\subsubsection{Logical gates}

Depending on the QEC code, some logical gates are transversal. That
is, they can be implemented in a bitwise fashion \cite{Nielsen2000},
i.e.,\ by mere local operations in the case of logical gates which
act only on a single logical qubit. In the five qubit code, the logical
$\sigma_{x}$, $\sigma_{y}$, and $\sigma_{z}$ operations are of
such kind, while the logical Hadamard gate $\frac{1}{\sqrt{2}}\left(\begin{array}{rr}
1 & 1\\
1 & -1
\end{array}\right)$ is not. In the Steane code on the other hand, the logical Hadamard
gate and even the logical CNOT gate are transversal, as well (next
to the logical $\sigma_{x}$, $\sigma_{y}$, and $\sigma_{z}$ operations)
\cite{Nielsen2000}. Unfortunately, neither the logical $\frac{\pi}{8}$
gate $\left(\begin{array}{cc}
1 & 1\\
1 & e^{i\frac{\pi}{4}}
\end{array}\right)$ is transversal nor any other gate which could complement the transversal
gates of the Steane code to a universal set of logical gates.

Therefore, we have chosen the logical Hadamard gate for the five qubit
code and the logical $\frac{\pi}{8}$ gate for the Steane code as
non trivial examples for logical gates. The presented sequences do
not amplify any correctable error present on the logical qubit (Sec.~\ref{sub:Operations-on-logical qubits}).
Abandoning this property would result in much simpler sequences for
the logical gates.

\begin{figure}
\includegraphics[width=0.767\columnwidth]{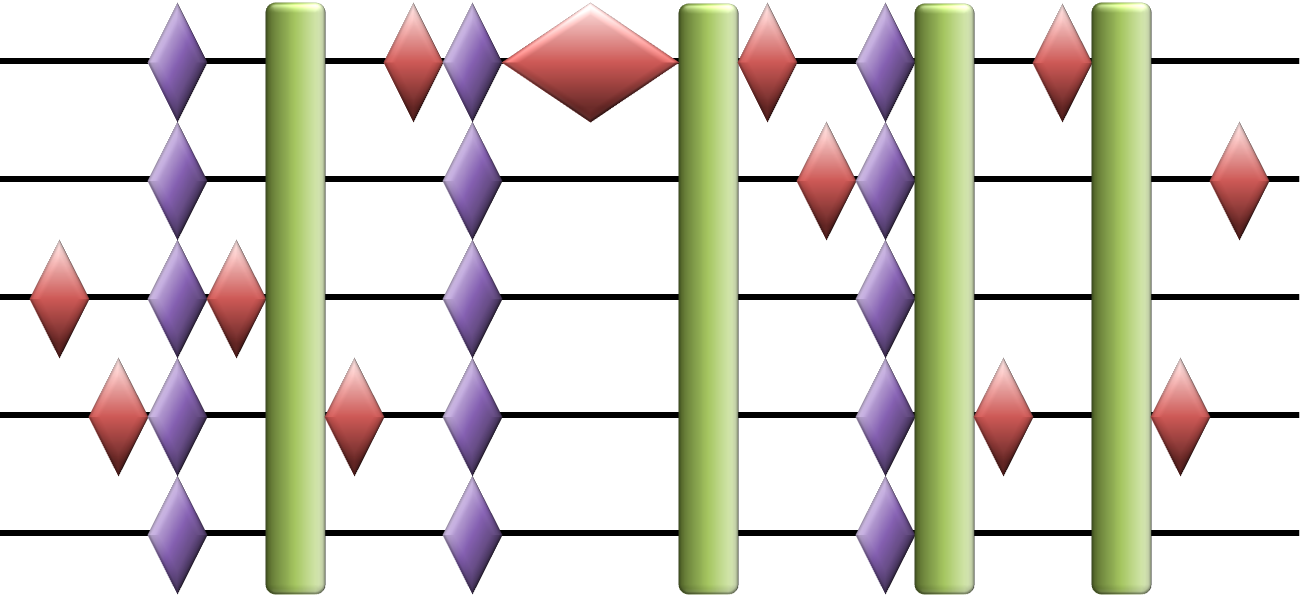}

\protect\caption{\label{fig:Hadamard 5 qubit code}Logical Hadamard gate for the five
qubit code \eqref{eq:Sequence Hadamard five qubit code}.}
\end{figure}

\paragraph{For the five qubit code, }

the logical Hadamard gate $\frac{1}{\sqrt{2}}\left(\begin{array}{rr}
1 & 1\\
1 & -1
\end{array}\right)$ (Fig.~\ref{fig:Hadamard 5 qubit code}) can be implemented by the
following sequence:
\begin{equation}
\begin{array}{lllll}
z_{3}\left(\frac{\pi}{2}\right) & z_{4}\left(\frac{\pi}{2}\right) & X\left(\frac{\pi}{2}\right) & z_{3}\left(\frac{\pi}{2}\right) & Y^{2}\left(\frac{\pi}{2}\right)\\
z_{4}\left(\frac{\pi}{2}\right) & z_{1}\left(\frac{\pi}{2}\right) & X\left(\frac{\pi}{2}\right) & z_{1}\left(-\frac{\pi}{2}\right) & Y^{2}\left(\frac{\pi}{2}\right)\\
z_{1}\left(\frac{\pi}{2}\right) & z_{2}\left(\frac{\pi}{2}\right) & X\left(\frac{\pi}{2}\right) & Y^{2}\left(\frac{\pi}{2}\right) & z_{4}\left(\frac{\pi}{2}\right)\\
z_{1}\left(\frac{\pi}{2}\right) & Y^{2}\left(\frac{\pi}{2}\right) & z_{4}\left(\frac{\pi}{2}\right) & z_{2}\left(\frac{\pi}{2}\right).
\end{array}\label{eq:Sequence Hadamard five qubit code}
\end{equation}

\begin{figure}
\includegraphics[width=0.785\columnwidth]{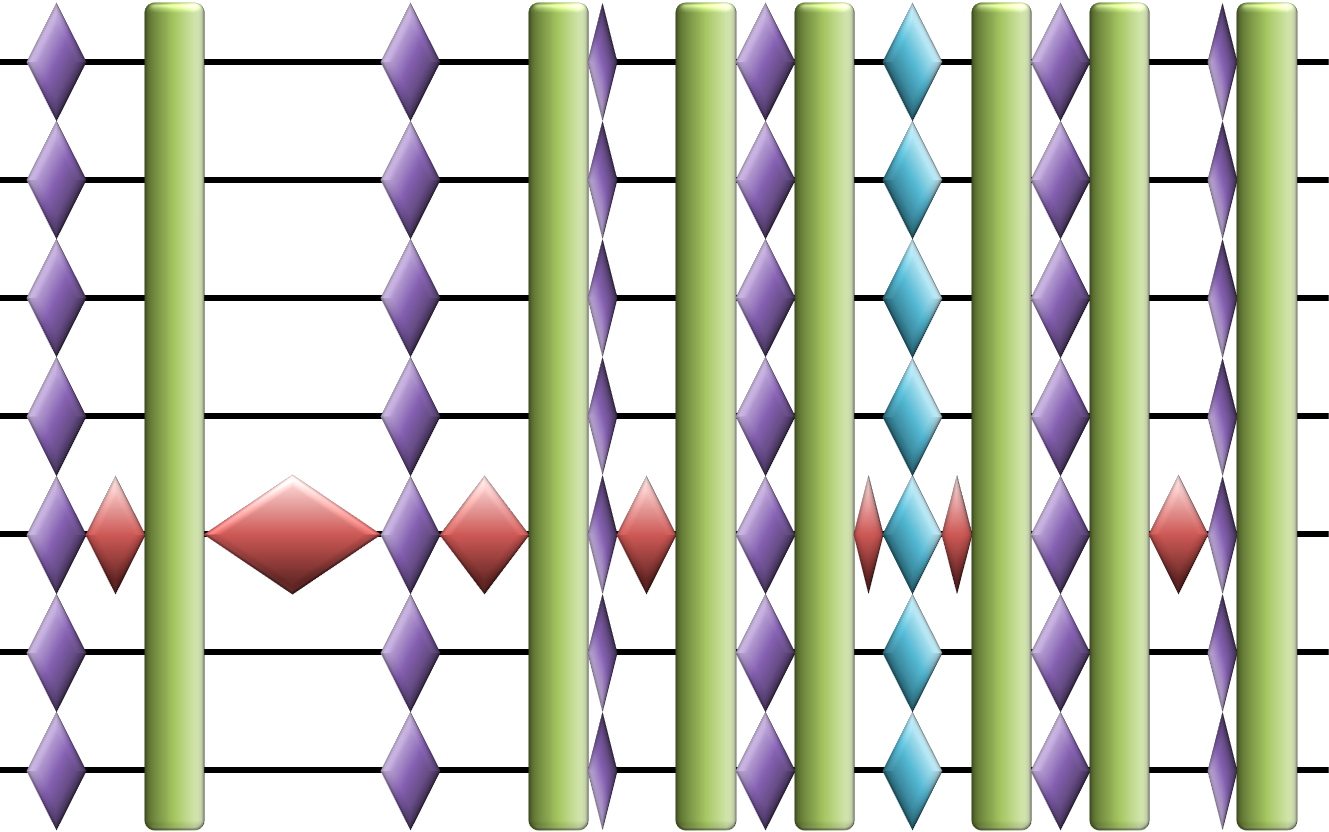}\protect\caption{\label{fig:-Pi achtel Steane}Logical $\frac{\pi}{8}$ gate for the
Steane code \eqref{eq:Sequence Pi achtel Steane}.}

\end{figure}

\paragraph{For the Steane code,}

the logical $\frac{\pi}{8}$ gate $\left(\begin{array}{cc}
1 & 1\\
1 & e^{i\frac{\pi}{4}}
\end{array}\right)$ (Fig.~\ref{fig:-Pi achtel Steane}) can be realized via
\begin{equation}
\begin{array}{lllll}
X\left(\frac{\pi}{2}\right) & z_{5}\left(\frac{\pi}{2}\right) & Y^{2}\left(\frac{\pi}{2}\right) & z_{5}\left(-\frac{\pi}{2}\right) & X\left(\frac{\pi}{2}\right)\\
z_{5}\left(\frac{3}{4}\pi\right) & Y^{2}\left(\frac{\pi}{2}\right) & X\left(-\frac{\pi}{4}\right) & z_{5}\left(\frac{\pi}{2}\right) & Y^{2}\left(\frac{\pi}{2}\right)\\
X\left(\frac{\pi}{2}\right) & Y^{2}\left(\frac{\pi}{2}\right) & z_{5}\left(\frac{\pi}{4}\right) & Y\left(\frac{\pi}{2}\right) & z_{5}\left(\frac{\pi}{4}\right)\\
Y^{2}\left(\frac{\pi}{2}\right) & X\left(\frac{\pi}{2}\right) & Y^{2}\left(\frac{\pi}{2}\right) & z_{5}\left(\frac{\pi}{2}\right) & X\left(-\frac{\pi}{4}\right)\\
Y^{2}\left(\frac{\pi}{2}\right).
\end{array}\label{eq:Sequence Pi achtel Steane}
\end{equation}
 
We remark that we also searched for sequences with an open parameter
$\alpha$ for arbitrary logical rotations $U_{L}(\alpha)$ of the
type 
\begin{equation}
U_{L}(\alpha)=\left(\begin{array}{cc}
1 & 1\\
1 & e^{i\alpha}
\end{array}\right).\label{eq:Beliebige Drehung}
\end{equation}
For the five qubit code, this search had to be terminated unsuccessful,
but for the Steane code, we managed to find such sequences for variable
$U_{L}(\alpha)$. While the sequences for the state preparation with
an open parameter $\alpha$ contain just one operation with an adjustable
argument \eqref{eq:WinkelZustand 5 Qubit} and \eqref{eq:Sequenz Stean Winkel},
the best sequences found for $U_{L}(\alpha)$ come with four such
operations. Of these four operations, only one shows a linear correspondence
between its argument and the parameter $\alpha$. For the other three,
we were not able to determine an analytical relation, which would
have allowed us to communicate these findings as a reproducible formula.

\section{Conclusion}

In conclusion, we have studied the gauge freedoms inherent in quantum
algorithms used for quantum error correction codes as well as in logical
operations on qubits in error protected states. We showed how these
gauge freedoms translate into specially chosen performance functions
suitable for optimal control algorithms. This allowed us to improve
the decomposition of the corresponding quantum algorithms into sequences
of elementary operations. These general ideas are complemented by
an in-detail description of an example algorithm tailored for experiments
with trapped ions (App.~\ref{sec:Optimization-algorithm-for-Trapped ions}).
As a result, various decompositions were presented, comprising quantum
algorithms for the three qubit, the five qubit, and the Steane code.
All these decompositions are exact and allow intuitive representation
as quantum circuits. The unitaries used for the decompositions can
all be implemented with fairly high fidelities on trapped ions such
that one might be optimistic to see experiments which realize some
of the more advanced quantum error correction codes in the near future.

\section{Acknowledgments}

This research was funded by the Austrian Science Fund (FWF): P24273-N16,
SFB F40-FoQus F4012-N16 and by the Austrian Ministry of Science BMWF
as part of the UniInfrastrukturprogramm of the Research Platform Scientific
Computing at the University of Innsbruck. I like to thank the group
of Prof.\ Rainer Blatt for their cooperation and the insights into
the experimental aspects of quantum computation, with special thanks
(in alphabetic order) to Thomas Monz, Daniel Nigg, Christian Roos
and Philipp Schindler. Further, I like to thank my supervisor Wolfgang
Dür for the freedom to pursue my own ideas.

I'm currently looking for a PostDoc position.

\bibliographystyle{revcompchem}
\bibliography{Zitate_OptimalQEC}

\appendix

\section{Elementary operations for trapped ions\label{sec:Elementary-operations-for-Trapped-ions}}

In this section, we provide a short explanation for the choice of
the effective unitaries $u_{t}(\alpha_{t})$ made in Eq.~\ref{eq:Effective Ham}.
We start with the Mølmer-Sørensen (MS) gates $\exp(-i\cdot\alpha_{t}\cdot S_{x/y}^{2})$,
which were chosen because they are currently the entangling gates
with the highest fidelities \cite{Benhelm2008}. They entangle each
ion with each other ion in the trap. A strong point of the MS gate
is that up to second order in perturbation theory it is independent
of the motional mode of the ions, while, e.g.,\ the formerly used
Cirac-Zoller CNOT gate \cite{Cirac1995} explicitly relies on the
assumption that the ions are in a well defined motional mode. More
details can be found in Ref.~\cite{Roos2008}.

For the MS gates, a bichromatic laser field is needed. Restricting
to a single frequency instead of the bichromatic field, the same laser
beams can be used for the collective rotations $\exp(-i\cdot\alpha_{t}\cdot S_{x/y})$.
Further, we need some local operations to ``individualize'' the
qubits. In the setup we have in mind, this requires an extra laser
beam which can be focused on single ions only. If this requirement
is not fulfilled with 100\% accuracy, the neighboring ions are affected
by the laser beam, as well. One advantage of the chosen single qubit
phase-shift gate $\exp(-i\cdot\alpha_{t}\cdot\sigma_{z}^{(j)})$ is
that it can be obtained from an ac Stark shift induced by an off-resonant
laser beam. The phase shift is proportional to the intensity of the
laser field and hence proportional to the square of the field amplitude.
Therefore, the residual effects that the laser beam exerts on the
neighboring ions drop off much faster than for operations which are
only linear in the field amplitude, as would be the case for $\exp(-i\cdot\alpha_{t}\cdot\sigma_{x/y}^{(j)})$.
As a further experimental advantage, the phase of the off-resonant
laser field does not need to be synchronized with the field used for
the MS gates $\exp(-i\cdot\alpha_{t}\cdot S_{x/y}^{2})$ and the collective
rotation $\exp(-i\cdot\alpha_{t}\cdot S_{x/y})$.

Another interesting option is to decouple ions \cite{Schindler2013},
i.e.,\ to transfer them into a state which is not affected by the
laser beams used for the gate operations. Although this might simplify
some operations, as, e.g.,\ in Eq.~\eqref{eq:Steane 3MS special},
decoupling ions entails some numerical drawbacks in the context of
an optimization algorithm. Either we have to increase the local Hilbert
spaces from dimension two to four to accommodate the decoupled states,
or we define, e.g.,\ extra MS gates which act only on a restricted
number of ions. In this case, due to the high number of combinatorial
possibilities to decouple different ions, the number of elementary
operations would strongly increase and most likely suffocate the optimization.
This argument is of course only true for a kind of black box optimization
with no physical insight, where all possible combinations of decoupled
ions are considered equally likely. If the structure of the problem
at hand clearly singles out a limited amount of useful MS gates which
only operate on a subset of all ions, it might be beneficial to enlarge
the set of elementary operations by these gates; see, e.g.,\  Ref.~\cite{Nigg2014}.
This is also true for Eq.~\eqref{eq:Steane 3MS special}, where an
extra MS gate operating on a special subset was included by hand.

\section{Optimization algorithm for trapped ions\label{sec:Optimization-algorithm-for-Trapped ions}}

In this section, we provide an in-detail description of the optimization
algorithm used for finding the results presented in Sec.~\ref{sec:Results}.

\subsection{Extra criteria\label{sub:Extra-criteria}}

So far, the only optimization objective discussed in Sec.~\ref{sec:Optimization}
is to maximize the performance function $\Phi(\boldsymbol{\alpha})$,
which ensures that the ansatz $\prod_{t}u_{t}(\alpha_{t})$ \eqref{eq:Prodult_Unitar}
mimics the desired target operation $U_{\textrm{target}}$ as well
as possible. For trapped ions, further criteria are of importance.
Due to the reduction to effective unitary operations, perfect solutions
are found much more frequently, which brings with it the need to define
a ranking of these otherwise perfect solutions.

\subsubsection{First criterion}

As primary criterion, we demand that the optimal sequence of the effective
unitary operations $\prod_{t=0}^{T}u_{t}(\alpha_{t})$ be short in
number (small $T)$. That is, we  have a few $u_{t}(\alpha_{t})$
with big values of $\alpha_{t}$ rather than a lot of $u_{t}(\alpha_{t})$
with small values of $\alpha_{t}$. Under this conditions, it might
seem that the arguments $\alpha_{t}$ are not the only unknown components
in the optimization problem, since we are neither a priori aware of
the minimal number of unitaries needed, nor do we know the order of
the different effective Hamiltonians $H_{j}$ we have to apply. 

The solution we adopt to find a short sequence of correctly ordered
unitaries is to start with a large (random) sequence $\prod u_{t}(\alpha_{t}^{\textrm{inital}})$
and gradually remove dispensable unitaries during the optimization
process. To gradually remove these unitaries, it suffices to include
side conditions which force most of the $\alpha_{t}$ to turn stepwise
into zero. These side conditions have to be adjusted individually
for each $\alpha_{t}$ such that they exert strong pressure on the
majority of unimportant $u_{t}(\alpha_{t})$, while the few important
$u_{t}(\alpha_{t})$ only be influenced marginally. Of course, we
can only judge the importance of a given unitary $u_{k}(\alpha_{k})$
at the actual moment in the optimization, which might differ from
its final importance at the end of the optimization. 

In Ref.~\cite{Nebendahl_Roos2009}, we derived the importance from
the length of $\alpha_{k}$. Here, we use an alternative method and
define the importance $I$ of a unitary $u_{k}(\alpha_{k})$ as the
difference 
\begin{equation}
I\left[u_{k}(\alpha_{k})\right]=\Phi(\boldsymbol{\alpha})-\Phi^{\bar{k}}(\boldsymbol{\alpha}),\label{eq:Importance}
\end{equation}
where $\Phi(\boldsymbol{\alpha})$ is the regular performance function
obtained from the actual sequence $\prod_{t=0}^{T}u_{t}(\alpha_{t})$,
while $\Phi^{\bar{k}}(\boldsymbol{\alpha})$ is obtained from the
same sequence without the unitary in question $u_{k}(\alpha_{k})$,
i.e.,\ $\prod_{t=0}^{k-1}u_{t}(\alpha_{t})\cdot\prod_{t=k+1}^{T}u_{t}(\alpha_{t})$.
In other words: A unitary $u_{k}(\alpha_{k})$ is unimportant if omitting
it completely has only a small impact on the performance function,
i.e.,\ $\Phi(\boldsymbol{\alpha})\approx\Phi^{\bar{k}}(\boldsymbol{\alpha})$.
Thanks to Eq.~\eqref{eq:Produkt aufspaltung}, the calculation of
$\Phi^{\bar{k}}(\boldsymbol{\alpha})$ is efficient.

\subsubsection{Second criterion\label{sub:Second-criterion}}

We like to introduce a second criterion for a good sequence $\prod u_{t}(\alpha_{t})$.
In the experimental realization, it is favorable to deal with only
one type of Mølmer-Sørensen (MS) gates $u_{\textrm{MS}}=\exp(-i\cdot\alpha_{\textrm{MS}}\cdot S_{x/y}^{2})$,
i.e.,\ we like $\alpha_{\textrm{MS}}$ to be the same for all MS
gates in the sequence. Further, the value of $\alpha_{\textrm{MS}}$
should be a natural fraction of $\pi$. Interestingly, this seems
to be a favorable numerical value, as well. In the best sequences
$\prod u_{t}(\alpha_{t})$ we found for various $U_{\textrm{target}}$,
all $\alpha_{t}$ (not just the $\alpha_{\textrm{MS}}$) turned out
to have quantized values of the type $\alpha_{t}=\frac{m_{t}}{2^{n_{t}}}\cdot\pi$,
with $m_{t}\in\mathbb{Z}$, $n_{t}\in\mathbb{N}_{0}$. Therefore,
including side conditions which favor such values also allows us to
improve the optimization process.

\subsection{Optimization}

The performance function might be optimized by various standard optimizers.
As already pointed out in Sec.~\ref{sub:Evaluating-the-performance},
the product structure of the performance function allows an efficient
calculation of its gradient such that one might choose a method such
as, e.g.,  GRAPE \cite{Khaneja2005}, which takes advantage of this
information. Still, we favored an alternative approach, which optimizes
all $\alpha_{t}$ individually based on a second-order Taylor expansion.
Due to its simple nature, this algorithm can be easily altered and
therefore allows a neat integration of the extra criteria formulated
in the last section, as we demonstrate below.

Except for trivial problems, the numerical maximization of the performance
function consists of many optimization cycles, where all $\alpha_{t}$
are adjusted repeatedly. To optimize one specific parameter $\alpha_{k}$,
all other parameters $\alpha_{0},\alpha_{1},\dots,\alpha_{k-1},\alpha_{k+1},\dots\alpha_{T}$
are kept constant mapping the multivariable function $\Phi(\boldsymbol{\alpha})$
to a single-variable function $\phi_{k}(\alpha_{k})$,
\begin{equation}
\Phi(\boldsymbol{\alpha})\equiv\Phi(\alpha_{0},\alpha_{1},\dots\alpha_{T})\rightarrow\phi_{k}(\alpha_{k}).
\end{equation}

The different parameters $\alpha_{k}$ are updated sequentially, starting
with $k=0$. Hereby, we take advantage of the method described in
Sec.~\ref{sub:Evaluating-the-performance}. To use Eq.~\eqref{eq:Produkt aufspaltung}
we need to know the expressions $\langle f_{lj}|\prod_{t=0}^{k-1}u_{t}(\alpha_{t})$
and $\prod_{t=k+1}^{T}u_{t}(\alpha_{t})|i_{lj}\rangle$ for each $k$.
The actual $\langle f_{lj}|\prod_{t=0}^{k-1}u_{t}(\alpha_{t})$ can
be derived from its predecessor $\langle f_{lj}|\prod_{t=0}^{k-2}u_{t}(\alpha_{t})$
after $\alpha_{k-1}$ has been updated, while all $\prod_{t=k+1}^{T}u_{t}(\alpha_{t})|i_{lj}\rangle$
are ideally calculated iteratively at the beginning of each optimization
cycle. 

With this preparation, we can use Eq.~\eqref{eq:Produkt aufspaltung}
and its obvious generalizations to compute $\phi_{k}$ and its first
and second derivative $\phi_{k}^{\text{\ensuremath{\prime}}},\phi_{k}^{\text{\ensuremath{\prime\prime}}}$
at the current value of $\alpha_{k}$. This allows us to approximate
$\phi_{k}$ by a parabola $\mathcal{P}_{k}$. The possibility to provide
individual parabola approximations $\mathcal{P}_{k}$ for all $\alpha_{k}$
without great effort is actually one of the nice features of this
simple optimization method.

If our sole objective was to maximize $\Phi(\boldsymbol{\alpha})$,
the obvious choice would be $\alpha_{k}^{[\textrm{new}]}=\alpha_{k}^{[\max]}$,
where $\alpha_{k}^{[\max]}$ is the position of the maximum of the
parabola, 
\begin{equation}
\mathcal{P}_{k}(\alpha_{k}^{[\max]})=\max\left(\mathcal{P}_{k}(x)\right).
\end{equation}
This should be backed up by some security protocols in case $\mathcal{P}_{k}$
turns out to have positive curvature, $\max\left(\mathcal{P}_{k}(x)\right)=\infty$
or the estimated function value is far off the real value $|\phi_{k}(\alpha_{k}^{[\textrm{new}]})-\mathcal{P}_{k}(\alpha_{k}^{[\textrm{new}]})|\gg\varepsilon$
. Since we also have to keep the criteria in mind that we formulated
in the last Sec.~\ref{sub:Extra-criteria}, we introduce a slightly
altered procedure.

\subsubsection{First criterion}

The first criterion states the preference of solutions where as many
$\alpha_{k}$ as possible have turned into zero. Since we do not possess
a priori knowledge which or how many $\alpha_{k}$ should become zero,
this criterion can not be phrased as an equation which has to be fulfilled
during the entire optimization. 

A feasible way is to add an adaptable potential $\Lambda(\boldsymbol{\alpha})$
to the performance function and maximize the sum $\Phi(\boldsymbol{\alpha})+\Lambda(\boldsymbol{\alpha})$.
The potential $\Lambda(\boldsymbol{\alpha})$ has to be designed such
that it drags the $\alpha_{k}$ of the less important unitaries $u_{k}(\alpha_{k})$
towards zero. In this fashion, the few important $u_{k}(\alpha_{k})$
are urged to compensate for the less important ones and finally replace
them. At the end of the entire optimization procedure, when only few
$\alpha_{k}>0$ have survived, $\Lambda(\boldsymbol{\alpha})$ must
be set to zero such that a pure maximum of $\Phi(\boldsymbol{\alpha})$
can be reached.

Instead of designing potentials $\Lambda(\boldsymbol{\alpha})$, we
can also go a much more direct way and simply set
\begin{eqnarray}
\alpha_{k}^{[\textrm{new}]} & = & \alpha_{k}^{[\max]}-\delta_{k}^{[0]}\label{eq:a_new}\\
 & \textrm{with} & |\delta_{k}^{[0]}|\leq|\alpha_{k}^{[\max]}|;\;\textrm{sign(\ensuremath{\alpha_{k}^{[\max]}}})=\textrm{sign(\ensuremath{\delta_{k}^{[0]}}}).\nonumber 
\end{eqnarray}
Obviously, $\delta_{k}^{[0]}$ pushes $\alpha_{k}^{[\textrm{new}]}$
closer towards zero but also prevents the sequence from fully maximizing
the performance function $\Phi$. Therefore, the $\delta_{k}^{[0]}$
have to be adopted during the optimization process, as described for
the potential $\Lambda(\boldsymbol{\alpha})$. So, what is the correct
size of the $\delta_{k}^{[0]}$? First, it should be noted that for
each optimization step the omitted improvement of the performance
function due to the displacement $\delta_{k}^{[0]}$ can be obtained
as
\begin{eqnarray}
\Delta_{k}^{[0]}\Phi(\delta_{k}^{[0]}) & = & \phi_{k}(\alpha_{k}^{[\max]})-\phi_{k}(\alpha_{k}^{[\textrm{new}]})\nonumber \\
 & = & \phi_{k}(\alpha_{k}^{[\max]})-\phi_{k}(\alpha_{k}^{[\max]}-\delta_{k}^{[0]}).
\end{eqnarray}
The value of the omitted improvement $\Delta_{k}^{[0]}\Phi(\delta_{k}^{[0]})$
can also be estimated with the help of the approximation parabola
$\mathcal{P}_{k}$
\begin{equation}
\Delta_{k}^{[0]}\Phi(\delta_{k}^{[0]})\approx\mathcal{P}_{k}(\alpha_{k}^{[\max]})-\mathcal{P}_{k}(\alpha_{k}^{[\max]}-\delta_{k}^{[0]}).\label{eq:Omitted improvement}
\end{equation}
The decisive advantage of this approximation is that it is easy to
invert. This allows us to proceed as follows: We decide which omitted
improvement $\Delta_{k}^{[0]}\Phi(\delta_{k}^{[0]})$ we are willing
to tolerate and with this value we directly infer the corresponding
displacement $\delta_{k}^{[0]}$. In case we get $|\delta_{k}^{[0]}|\geq|\alpha_{k}^{[\max]}|$,
we include the (non-analytic) rule to set $\alpha_{k}^{[\textrm{new}]}=0$
and erase the unitary $u_{k}(\alpha_{k})$ from the sequence.

Hence, the new question arises: What do we choose as tolerable $\Delta_{k}^{[0]}\Phi$?
Following Sec.~\ref{sub:Extra-criteria}, $\Delta_{k}^{[0]}\Phi$
should be chosen as a reciprocal function of the importance $I[u_{k}(\alpha_{k}^{[\max]})]$
\eqref{eq:Importance}. Just to give an example (not as exclusive
choice), in our calculations we frequently used 
\begin{equation}
\Delta_{k}^{[0]}\Phi=\gamma^{[0]}\cdot\left[\left(\frac{0.25}{I[u_{k}(\alpha_{k}^{[\max]})]}\right)^{5}+1\right].\label{eq:Inverse Importance}
\end{equation}
The coupling factor $\gamma^{[0]}$ is adapted according to the stage
of optimization: At the very beginning, the system needs some optimization
time to get close enough to the maximum to develop a preliminary hierarchy
of important and unimportant unitaries in the sequence $\prod_{t}u_{t}(\alpha_{t})$.
Hence, $\gamma^{[0]}$ starts very low in the beginning and gradually
increases during the optimization. This increase often comprises many
orders of magnitude. Only to the very end, when all unimportant unitaries
have died away, $\gamma^{[0]}$ is set to zero to reach an undisturbed
maximum of the performance function $\Phi.$

The value of the coupling constant $\gamma^{[0]}$ directly influences
the number of unitaries in the sequence. If this number is reduced
too quickly, there might not be sufficient degrees of freedom left
to reach the maximum, while too many unimportant unitaries in the
sequence might prove fatal as well, since they tend to suffocate the
optimization. Any protocol used to adjust $\gamma^{[0]}$ should find
the right balance between these two situations.

\subsubsection{Second criterion}

The second criterion formulated in Sec.~\ref{sub:Extra-criteria}
favors quantized values $\alpha_{k}=\frac{m_{k}}{2^{n_{k}}}\cdot\pi$,
with $m_{k}\in\mathbb{Z}$, $n_{k}\in\mathbb{N}_{0}$. With slight
modification, this criterion can be enforced with a similar method
as used to drag the $\alpha_{k}$ towards zero. First, we replace
$\delta_{k}^{[0]}$ in Eq.~\eqref{eq:a_new} by a new displacement
$\delta_{k}^{[\textrm{quant}]}$ which pushes $\alpha_{k}^{[\textrm{new}]}$
towards the quantized value $\frac{m_{k}}{2^{n_{k}}}\pi$ which is
next to $\alpha_{k}^{[\max]}$. The replacement of $\delta_{k}^{[0]}$
by $\delta_{k}^{[\textrm{quant}]}$ is most usefully performed towards
the end of the optimization process, when $\delta_{k}^{[0]}$ has
done most of its job and the number of unitaries in the sequence is
already strongly reduced. Analog to $\delta_{k}^{[0]}$, the displacement
$\delta_{k}^{[\textrm{quant}]}$ is derived from a pre-defined $\Delta_{k}^{[\textrm{quant}]}\Phi$
\eqref{eq:Omitted improvement}. While the strength of $\Delta_{k}^{[0]}\Phi$
is calculated using the importance $I[u_{k}(\alpha_{k}^{[\max]})]$
\eqref{eq:Inverse Importance}, there is no particular reason to do
so for $\Delta_{k}^{[\textrm{quant}]}\Phi$. It seems more meaningful
to relate the value of $\Delta_{k}^{[\textrm{quant}]}\Phi$ to the
proximity of $\alpha_{k}^{[\max]}$ to the next quantized value. That
is, $\Delta_{k}^{[\textrm{quant}]}\Phi$ is set to zero if $\alpha_{k}^{[\max]}$
is placed in the middle between two quantized values and increases
when $\alpha_{k}^{[\max]}$ approaches one of them. 

To avoid false expectations, it should be mentioned that finding solutions
where all $\alpha_{k}$ are nicely quantized is supported by the described
method but not guaranteed. Using a multi-ansatz with many different
initial sequences is still a key element to success. 

For the Mølmer-Sørensen (MS) gates, we actually use a far more drastic
approach. Since we favor solutions with only one type of MS gates
$u_{\textrm{MS}}=\exp(-i\cdot\alpha_{\textrm{MS}}\cdot S_{x/y}^{2})$,
it turned out to be advantageous to fix their number and their $\alpha_{\textrm{MS}}$
already in the initial sequence $\prod u_{t}(\alpha_{t}^{\textrm{inital}})$
and never allow the computer to change these values during the optimization.
Of course, this obliges us to optimize over different types of initial
sequences, but we found that the improvements gained in the optimization
compensate by far for this little inconvenience.

\subsubsection{Simulated annealing \label{sub:Simulated-annealing}}

Numerical algorithms based on stepwise local optimization often face
the problem that they get stuck in local extrema. Algorithms built
on simulated annealing \cite{Kirkpatrick1983} try to avoid this problem
by accepting locally suboptimal moves too, albeit with a reduced probability.
To incorporate an analog effect, one might introduce a third displacement
$\delta_{k}^{[\textrm{sim. anneal.}]}$ stemming from $\Delta_{k}^{[\textrm{sim. anneal.}]}\Phi$
which is randomly drawn from an adequate Boltzmann distribution. By
adapting $\Delta_{k}^{[\textrm{sim. anneal.}]}\Phi$, we can tune
the algorithm continuously between an optimal local search and simulated
annealing. Further, this algorithm never does an optimization in vain,
contrary to ``classical'' simulated annealing, where certain results
are rejected with a probability depending on the outcome.

Alternatively, the system can be disturbed and pushed out of a local
extremum by a variation of the coupling factors $\gamma^{[0]}$ and
$\gamma^{[\textrm{quant}]}$. Probably the most powerful method to
escape a local trap is to complement the sequence $\prod u_{t}(\alpha_{t})$
with new unitaries $u_{\textrm{new}}(\alpha\approx0)$ close to the
identity. Of course, this method must be handled wisely, since it
counteracts the first criterion, which favors short sequences.

\subsection{Protocols}

The optimization algorithm described above contains the open parameters
$\gamma^{[0]}$ \eqref{eq:Inverse Importance} and $\gamma^{[\textrm{quant}]}$,
$\gamma^{[\textrm{sim. anneal.}]}$ (not explicitly mentioned) to
adjust $\Delta_{k}^{[0]}\Phi$, $\Delta_{k}^{[\textrm{quant}]}\Phi$,
and $\Delta_{k}^{[\textrm{sim. anneal.}]}\Phi$. Hence, we need a
protocol telling us how these parameters should be adopted according
to the stage of the optimization process. Unfortunately, it is neither
straightforward to fathom the stages nor to associate them with precise
values for the parameters. We spent a considerable amount of time
to develop a good heuristic which does this job, but most of it is
based on intuition and only few on scientific facts. Therefore, we
do not deem it appropriate to go into any greater detail here.

Still, one further optimization protocol is worth mentioning: The
optimization routine we have described so far starts with a random
sequence and tries to end up with a sequence $\prod_{t}u_{t}(\alpha_{t})$
which reproduces the target operation $U_{\textrm{target}}$ (up to
a gauge transformation). Once we have found such a sequence and saved
it to the disk, we might wish to go on optimizing it and, e.g.,\ try
to find a shorter solution. Here, the problem is that each solution
is already a maximum of the performance function and therefore the
optimization gets stuck in a local trap.

\subsubsection{Disturbing the sequence\label{sub:Disturbing-the-sequence}}

To escape such a trap, the algorithm has to be disturbed. While the
methods described in Sec.~\prettyref{sub:Simulated-annealing} tend
to disturb the sequence everywhere a little bit, we found it more
promising to introduce a few strong local disturbances. The motivation
for this approach roots in the observation that improvements \---
even those obtained with other means \--- often affect only very
few unitaries in the sequence and leave the rest unchanged. 

In concrete terms, the procedure looks as follows: We start with the
perfect sequence $\prod_{t}u_{t}(\alpha_{t})$ and choose one unitary
$u_{k}(\alpha_{k})$ by random. This unitary is either replaced by
$u_{k}(\alpha_{k})\rightarrow u_{k}(-\alpha_{k})$ or by $u_{k}(\alpha_{k})\rightarrow u_{k}(0)$
(choice by random). If the algorithm is able to restore the sequence,
we repeat the procedure with the same disturbance plus one more. This
is done until the algorithm fails to repair the damage. Then, the
sequence is filled up with (lots of) fresh unitaries $u_{\textrm{new}}(\alpha\approx0)$
and optimized once more. If no improvement has been found until now,
we start again with the perfect sequence $\prod_{t}u_{t}(\alpha_{t})$
and a new single disturbance.

We like to point out that this approach might become of greater importance
in future applications, when quantum algorithms reach a complexity
where solutions can no longer be found by ab initio methods. In this
case, solutions have to be assembled from solutions of solvable subproblems.
These assembled solutions are also maxima of the performance function
but might still be optimizable with the procedure described above.
An example for such an approach can be found in the Results Sec.~\prettyref{sub:Stabilizer},
where the assembled solution of the joint measurement of all four
stabilizers of the five qubit code allowed for quite a successful
optimization, which reduced the original assembled solution consisting
of $52$ unitaries to a solution comprising only $30$ unitaries.
\end{document}